\documentclass[11pt]{article}
\usepackage{jheppubmod}
\pdfoutput=1
\usepackage[showonlyrefs]{mathtools}
\usepackage{psfrag}
\usepackage{array}
\usepackage{amssymb}
\usepackage{amsthm}
\usepackage{centernot}
\usepackage{graphicx}
\usepackage{wrapfig}
\usepackage{slashed}
\usepackage[labelsep=quad]{subcaption}
\mathtoolsset{showonlyrefs}
\usepackage{epstopdf}
	
\usepackage{epsfig}
\usepackage[punctsep]{collref}
\collectsep[]{;}

\def\alrcoarse{{\cal A}_{\regcoarse}}
\def\alrfine{{\cal A}_{\regfine}}
\def\alr{{\cal A}_{\reg}}
\def\albcoarse{{\cal A}_{\bcoarse}}
\def\albfine{{\cal A}_{\bfine}}

\def\al{A}
\def\alher{\al}
\def\bl{B}
\def\idmat{{1}}

\def\ha{{\cal H}}
\def\hab{\overline{\cal H}}
\def\st{\psi}
\def\stlocc{\locc(\st)}
\def\sstlocc{\locc(\sst)}

\def\locc{{\cal L}}
\def\loccstate{L}
\def\loccn{n_{m}}
\def\ncut{N_{c}}
\def\combop{{\cal X}}
\def\sstcombop{{\cal Y}}
\def\spect{{\cal SP}}
\def\onest{{\bf 1}}

\def\sst{\phi}
\def\pst{\Psi}
\def\psst{\Phi}

\def\hst{{\cal H}_{\st}}
\def\hsst{{\cal H}_{\sst}}

\def\tr{{\rm Tr}}

\def\Sep{{\cal D}_{\text{sep}}}
\def\Spure{{\cal D}_{\text{pure}}}

\def\Or[#1]{{\text{O}}\left({#1}\right)}
\def\dotl[#1,#2]{\left\langle #1,\, #2 \right\rangle}
\def\dotlb[#1,#2]{\left\langle #1,\, #2 \right\rangle}
\def\dotlm[#1,#2]{\left[ #1,\, #2 \right]}
\def\dotp[#1,#2]{(\vect{#1} \cdot\vect{#2})}
\def\aff[#1,#2]{\hat{#1}(#2)}
\def\n4sym{{\cal N}=4 SYM}
\def\>{\rangle}
\def\<{\langle}
\def\weight[#1,#2,#3]{\{(#1),#2,#3\}}
\def\ads[#1]{$\text{AdS}_{#1}$}
\def\cft[#1]{$\text{CFT}_{#1}$}

\newcommand{\bcoarse}{{\cal B}_{\text{C}}}
\newcommand{\regcoarse}{{\reg}_{\text{C}}}
\newcommand{\regfine}{\reg_{\text{F}}}
\newcommand{\diam}{{\cal D}}
\newcommand{\bfine}{{\cal B}_{\text{F}}}
\newcommand{\bfinefine}{\cal B}
\newcommand{\cauch}{S}
\newcommand{\hamilt}[1]{{\cal H}\{#1\}}
\newcommand{\causalcomp}[1]{\widehat{#1}}

\newcommand{\ent}[2]{{E_{#1}(#2)}}
\newcommand{\modop}[1]{{\Delta_{#1}}}

\newcommand{\relmod}[2]{\Delta(#1 | #2)}
\newcommand{\relmodsmall}[3]{\Delta_{#1}(#2 | #3)}
\newcommand{\rels}[2]{{\cal S}_{(#1|#2)}}
\newcommand{\relsd}[2]{{\cal S}^{\dagger}_{(#1|#2)}}
\newcommand{\snorm}[3]{S^{\|}_{#1}{(#2|#3)}}
\newcommand{\chinorm}[3]{\chi^{\|}_{#1}{(#2|#3)}}
\newcommand{\gendist}[3]{D_{#1}{(#2,#3)}}
\newcommand{\distfun}{D}

\newcommand{\relent}[3]{S^{\text{ar}}_{#1}(#2|#3)}

\newcommand{\relentnew}[3]{\tilde{S}^{\text{ar}}_{#1}(#2|#3)}
\renewcommand{\S}[1]{{\cal S}_{#1}}
\newcommand{\op}{{\cal O}}

\newcommand{\alset}{{\cal A}}
\newcommand{\blset}{{\cal B}}

\newtheorem*{assumption}{Special Case \label{h:sc}{(SC)}}
\newcommand{\specialcase}{{\hyperref[h:sc]{SC}}}

\newcommand{\be}{\begin{equation}}
\newcommand{\ee}{\end{equation}}
\newcommand{\ba}{\begin{align}}
\newcommand{\ea}{\end{align}}
\newcommand{\bs}{\begin{split}}

\def\sess\end{split}

\newcommand{\vect}[1]{{\vec{#1}}}
\newcommand{\caus}[1]{{\cal C}_{#1}}

\def\reg{{\cal R}}
\def\bulkreg{{\cal B}}
\def\dima{{\cal D}}

\title{Quantum information measures for restricted sets of observables}
\author{Sudip Ghosh and Suvrat Raju}
\affiliation{International Centre for Theoretical Sciences, Tata Institute of Fundamental Research, Shivakote, Bengaluru 560089, India.}
\emailAdd{sudip.ghosh@icts.res.in}
\emailAdd{suvrat@icts.res.in}

\date{}

\abstract{
We study measures of quantum information when the space spanned by the set of accessible observables is not closed under products, i.e., we consider systems where an observer may be able to measure the expectation values of two operators, $\langle O_1 \rangle $ and  $\langle O_2 \rangle$, but  may not have access to $\langle O_1 O_2 \rangle$. This problem is relevant for the study of localized quantum information in gravity since the set of approximately-local operators in a region may not be closed under arbitrary products.  While we cannot naturally associate a density matrix with a state in this setting, it is still possible to define a modular operator for a state, and distinguish between two states using a relative modular operator.  These operators are defined on a ``little Hilbert space'', which parameterizes small deformations of the system away from its original state, and they do not depend on the structure of the full Hilbert space of the theory.  We extract a class of relative-entropy-like quantities from the spectrum of these operators that  measure the distance between states, are monotonic under contractions of the set of available observables, and vanish only when the states are equal. Consequently, these distance-measures can be used to define measures of bipartite and multipartite entanglement.  We describe applications of our measures to ``coarse-grained'' and ``fine-grained'' subregion dualities in AdS/CFT and provide a few sample calculations to illustrate our formalism.}
\begin{document}
\maketitle
\section{Introduction and Summary of Results}
In studying entanglement and other quantum information measures associated with a system, we often assume that the Hilbert space factorizes as $\ha \otimes \hab$ with a factor, $\ha$ associated with the system of interest and another factor, $\hab$ associated with the rest of the world. A slightly more general way to state this assumption is that we assume that we have access to an {\em algebra} of observables that characterizes the system of interest. In the case where the Hilbert space admits a bipartite decomposition, this algebra simply comprises the set of all operators  in the theory that act trivially on $\hab$ but may act non-trivially on $\ha$. This set of operators forms a linear space and is closed under multiplication and the adjoint operation.

However, it often happens that while we can measure the expectation value of one operator in the system, say $\langle O_1 \rangle$, and also another operator, say $\langle O_2 \rangle$, we {\em cannot} measure the product $\langle O_1 O_2 \rangle$ due to some physical constraints. In this paper, we explore the extent to which one can define various quantum information measures in such settings --- when the space spanned by the available observables is not an algebra because it is not closed under multiplication.

We were originally motivated to study this problem with a view to understanding local entanglement in theories of quantum gravity. In theories of quantum gravity, except for some special regions,  the set of approximately-local operators associated with a region does not form an algebra  \cite{Ghosh:2016fvm, Ghosh:2017pel,Banerjee:2016mhh}.

This point can be understood by contrasting  gravity with local quantum field theories. In a local quantum field theory, given a region, $\reg$,  the product of any two local operators in $\reg$ is another operator in $\reg$ and therefore the set of all local operators that belong to $\reg$ is closed under multiplication. This algebraic structure holds even in gauge theories, where the algebras associated with regions contain a center \cite{Casini:2013rba,Soni:2015yga,Ghosh:2015iwa}. The presence of this center creates ambiguities in quantum information measures, since the center can be considered either to belong to the region or to belong to its complement. But this ambiguity does not alter the fact that the set of local operators is closed under multiplication.

However, in a theory of quantum gravity, the situation is more subtle. Although there are no exactly local gauge invariant operators in gravity, it is possible, in a sense, to approximately localize simple operator in a region $\reg$. However, if we start considering sufficiently complicated polynomials of these operators, then (except for some special regions) these complicated polynomials do not remain confined to $\reg$ in any meaningful sense. Therefore the set of approximately local observables  does {\em not} span an algebra in a theory of quantum gravity.  As a corollary,  the Hilbert space of the theory does {\em not} factorize into a Hilbert space associated with $\reg$ and another factor associated with its complement. An explicit example of this phenomenon was given in \cite{Banerjee:2016mhh} motivated by the results of \cite{Papadodimas:2015jra, Papadodimas:2015xma, Papadodimas:2013jku,Papadodimas:2013wnh}, where the lack of factorization of the Hilbert space was important for the resolution of the information paradox proposed there. We give some additional examples in section \ref{secadscft}.

In the absence of an algebraic structure,  there is no natural way to associate a  density matrix with $\reg$. However, as we will describe below, it is still possible to define a related quantity, called a {\em modular operator}.  We can obtain some intuition for this object by thinking of  a special case. Consider an ordinary local quantum field theory without gauge fields, where the Hilbert space factorizes as $\ha \otimes \hab$ and where, with respect to the full set of operators (including operators in $\hab$), the system is in a pure state. Then, if  the density matrix of $\reg$ is $\rho$, the density matrix of $\hab$ is $\overline{\rho}$ which has the same spectrum as $\rho$. In this setting, the modular operator associated with $\reg$ is  $\rho \otimes \overline{\rho}^{-1}$. We also study the  {\em relative modular operator}, which distinguishes between states. In the setting above, given two states, where the density matrix associated to $\reg$ is $\rho$ and $\sigma$ respectively, the relative modular operator between the states is $\sigma \otimes \overline{\rho}^{-1}$. 

In the absence of a direct-product factorization, we will show that the modular and the relative modular operator continue to exist even though they {\em cannot be interpreted} simply in terms of density matrices.

We use the modular and the relative modular operators to define measures of distance between states. As we review below, an appropriate  measure of distance underpins the study of other measures of quantum information, such as entanglement. In the usual setting, a common and useful measure of distance is provided by the relative entropy. While we are not able to directly generalize the relative entropy, we are able to find other measures that share the nice properties of the relative entropy.

We now describe our results in some more detail. We consider a set of operators, which we denote by $\alset$. We assume that this set forms a complex vector space and also that it is closed under the adjoint operation: $\al \in \alset \implies \al^{\dagger} \in \alset$.
For notational simplicity, we also take  $1 \in \alset$. We set $\dim(\alset) = \dima$.  We denote the state of the system by $\st$. We do not insist that $\st$ be pure, and this is reflected in our notation. We use $\st(\al) $ to denote expectation values of $\al$ in the state in $\st$. The action of the elements of $\alset$ on $\st$ creates a {\em little Hilbert space}, 
\be
\hst = \text{span} \{ |\al_1 \rangle, \ldots |\al_{\dima} \rangle \},
\ee
whose vectors are in one-to-one correspondence with elements of $\alset$ and where the norm is set by the state: $\langle \al_j | \al_i \rangle = \st(\al_j^{\dagger} \al_i)$. The little Hilbert space also contains a vector dual to the identity operator, $1$, denoted by $|\onest \rangle$.  Note that while the structure of the little Hilbert space requires us to measure one-point and two-point functions of operators in $\alset$, we will {\em not} require any higher point functions.

We then define information measures using constructs that operate {\em entirely within} this little Hilbert space. This is an important physical constraint in our analysis and reflects the fact that the observer may have no prior information about the structure of the full Hilbert space of the theory.

The modular operator on $\hst$ is defined as
\be
\langle \al_j | \modop{\st} | \al_i \rangle = \st(\al_i \al_j^{\dagger}).
\ee
The relative modular operator is also an operator within $\hst$. If we are given another state $\sst$, then the relative modular operator is defined as
\be
\langle \al_j | \relmod{\st}{\sst} | \al_i \rangle = \sst(\al_i \al_j^{\dagger}).
\ee

In the case where $\alset$ forms an algebra, Araki \cite{araki1976relative} defined the relative entropy as
\be
\label{defrelentar}
\relent{\alset}{\st}{\sst} = -\langle \onest | \log \big[\relmod{\st}{\sst} \big] | \onest \rangle.
\ee
This definition can be adopted, as it stands, to the case where $\alset$ is not an algebra. However, we find that the physical quantity, so defined, shares some but {\em not} all of the  desirable properties that the relative entropy has when $\alset$ is an algebra.

Even, in the case where $\alset$ is not closed under multiplication, we find that $\relent{\alset}{\st}{\sst}$ continues to be nonnegative. $\relent{\alset}{\st}{\sst}$ is also  monotonic under contractions of the set of available observables, i.e., if we consider a subset of observables whose span is a subspace of the original space, $\blset \subset \alset$, then $\relent{\blset}{\st}{\sst} \leq \relent{\alset}{\st}{\sst}$. However, it turns out that it is {\em no longer} true that this quantity vanishes only when the states are the same.  
\be
\relent{\alset}{\st}{\sst} = 0 \centernot \implies \st(\al_i \al_j) = \sst(\al_i \al_j).
\ee

Therefore, we introduce two additional measures of distance between states. Let $\combop = (\modop{\st})^{-{1 \over 2}} \relmod{\st}{\sst} (\modop{\st})^{-{1 \over 2}}$ and $\|\combop\|$, and  $\|\combop^{-1}\|$ denote the operator norms of $\combop$ and its inverse respectively. Then we set
\be
\label{normmeasintro}
\begin{split}
&\snorm{\alset}{\st}{\sst} = \log \|\combop \| + \log \|\combop^{-1}\|, \\
&\chinorm{\alset}{\st}{\sst}= 1 - f_{\chi}(\|\combop\|) - f_{\chi}(\|\combop^{-1}\|),
\end{split}
\ee
with $f_{\chi}(x) = {2 x \over (1 + x)^2}$. This choice of $f_{\chi}$ is not unique and we describe how the $\chi$-distance can be generalized in various ways.
Even though it is not obvious from the formulas above, both the measures in \eqref{normmeasintro} are {\em symmetric} between $\st$ and $\sst$.

We show that both  $\snorm{\alset}{\st}{\sst}$ are $\chinorm{\alset}{\st}{\sst}$ are nonnegative and they both have the property that
\be
\label{discernintro}
\chinorm{\alset}{\st}{\sst} = 0 \iff \snorm{\alset}{\st}{\sst} = 0 \iff \st(\al_i \al_j) = \sst(\al_i \al_j),
\ee 
i.e., when these quantities vanish, the states $\st, \sst$ are indistinguishable through the measurement of one- and two-point functions of $\alset$. Just like the relative entropy both these measures of distance are {\em monotonic} --- if we shrink $\alset$ by dropping some observables, the distance between states either remains constant or decreases. Moreover, both these measures  share an  additional important property of the relative entropy, which we call {\em insularity}. If we simply add on an ancillary system in a state $\st_{\text{anc}}$ whose operators, $\alset_{\text{anc}}$ are unentangled with the original system, then this does not change the distance between states: $\snorm{\alset \otimes \alset_{\text{anc}}}{\st \otimes \st_{\text{anc}}}{\sst \otimes \st_{\text{anc}}} = \snorm{\alset}{\st}{\sst}$ and $\chinorm{\alset \otimes \alset_{\text{anc}}}{\st \otimes \st_{\text{anc}}}{\sst \otimes \st_{\text{anc}}} = \chinorm{\alset}{\st}{\sst}$.
(The notation for product states, used in this relation, is standard but such states are defined more precisely in \eqref{prodstatedef}.)

$\snorm{\alset}{\st}{\sst}$ can be interpreted as an entropy, since it is {\em additive.} More precisely
\be
\snorm{\alset_1 \otimes \alset_2}{\st_1 \otimes \st_2}{\sst_1 \otimes \sst_2} = \snorm{\alset_1}{\st_1}{\sst_1} + \snorm{\alset_2}{\st_2}{\sst_2}.
\ee
$\snorm{\alset}{\st}{\sst}$ has the somewhat undesirable property that the distance between states can easily become infinite, as we explain in greater detail below. On the other hand, $\chinorm{\alset}{\st}{\sst}$ is not additive for product states, but it has the advantage that it always varies between $[0,1]$.

The measures $\snorm{\alset}{\st}{\sst}$ and $\chinorm{\alset}{\st}{\sst}$ are just two of a large class of distance measures, derived from  the spectrum of the modular and relative modular operators,   which can discern between states as in \eqref{discernintro}, and are monotonic and insular as described above. We denote any measure of distance that satisfies these properties through $\gendist{\alset}{\st}{\sst}$.

We describe how any such measure, $\gendist{\alset}{\st}{\sst}$,   can be used to define a measure of bipartite and multipartite entanglement. The key point is that notions of purity and separability can be generalized easily to our setting. A state, $\st$, is pure with respect to $\alset$ if it cannot be written as a convex combination of two other states: $\nexists \st_1, \st_2, 0 < \lambda < 1$ such that $\st = \lambda \st_1 + (1 - \lambda) \st_2$. Similarly, given a direct product decomposition,  $\alset = \alset_1 \otimes \alset_2$, we can define separable states as those that can be written as convex combinations of product states. A measure of bipartite entanglement is then given by
\be
\ent{\alset_1 \otimes \alset_2}{\st} = \underset{\sst \in \Sep}{\text{inf}} \gendist{\alset_1 \otimes \alset_2}{\st}{\sst},
\ee
where the infimum is taken over the set of all separable states. If the space of observables admits a {\em multipartite} factorization $\alset = \alset_1 \otimes \ldots \alset_n$ then the definition above immediately generalizes to a measure of multipartite entanglement. 

The virtue of this measure of entanglement is that, first, it measures only quantum and not classical correlations; second, it is invariant under local unitary transformations; and third that it remains constant or decreases under local operations and classical communication (LOCC).

We describe some applications of our information measures to  subregion duality in AdS/CFT \cite{Maldacena:1997re,Gubser:1998bc,Witten:1998qj}  --- the question of what region in the bulk is probed by a given spacetime region, $\reg$, on the boundary. It is generally believed that this bulk region is given by the so-called ``entanglement-wedge'' of the boundary region \cite{Headrick:2014cta}.  This duality can be most precisely phrased in terms of entanglement measures, and is equivalent to the claim that ``bulk relative entropy on the entanglement wedge is equal to boundary relative entropy in the region $\reg$'' \cite{Jafferis:2015del,Dong:2016eik}. This duality  is a statement about the bulk dual to  {\em all} operators in the region $\reg$ --- this is what one may call a ``fine grained'' subregion duality.

However, it is often convenient to consider another version of the subregion duality, which which we call a ``coarse grained'' duality.  Here,   we consider the set of simple operators in the region $\reg$, which we call $\alrcoarse$,  and likewise restrict bulk observables in the dual region to simple polynomials in local field operators, which we call $\albcoarse$. In this setting, we can also consider arbitrary regions $\reg$, which extend in both space and {\em time} and may {\em not} be causally complete. Then, we argue that the bulk region dual to $\reg$ in a coarse grained sense is given by
 \be 
\label{coarseadssub}
\bcoarse = \bigcup_{\diam \subseteq \reg} \caus{\diam},
\ee
which indicates the union of the {\em causal wedges}, $\caus{\diam}$ of each causal diamond, $\diam$ that belongs to $\reg$. (Causal wedges are defined in \eqref{causdef}.)

In the conventional formalism, it is difficult to write down an information-theoretic analogue of \eqref{coarseadssub} since the set of simple operators in $\reg$ do not form an algebra. However, in our formalism, if $\st$ and $\sst$ are two states then  \eqref{coarseadssub} simply implies
\be
\label{coarseadsent}
\gendist{\alrcoarse}{\st}{\sst} = \gendist{\albcoarse}{\st}{\sst},
\ee
where $\gendist{\alset}{\st}{\sst}$ is any one of the  measures of distance described above.

We also briefly comment on what happens when we consider the set of {\em all} operators in $\reg$ including arbitrarily complicated operators. Although we are not able to prove the entanglement-wedge proposal, we are able to show that, in this limit, the bulk dual, which we denote by $\bfinefine$ at least contains a region
\be
\label{finegrained}
\bfinefine \supseteq \bfine = \bigcup_{\cauch \subseteq \reg} \caus{\causalcomp{S}}
\ee
where the union runs over all spacelike intervals, $\cauch$, that lie in $\reg$ and $\caus{\causalcomp{\cauch}}$ indicates the causal wedge of the causal diamond built on $\cauch$. Our proof only uses principles from canonical gravity and, in particular, the fact that the Hamiltonian can be expressed as a boundary term. 

In terms of information measures, if $\alr$ is the set of all operators in $\reg$ and $\albfine$ is the set of all  simple bulk operators in $\bfine$ then \eqref{finegrained} implies that
\be
\label{fineadsent}
\gendist{\alr}{\st}{\sst}\geq  \gendist{\albfine}{\st}{\sst}.
\ee
The inequality arises because the region $\bfine$ is, in general, smaller than the entanglement wedge and because the measure of distance is monotonic.

After a description of our setup in section \ref{secsetup}, we turn to a detailed discussion of the modular and relative modular operator in section \ref{secmodrelmod}. We then describe how the spectrum
of these operators can be used to define measures of distance, and also entanglement, in section \ref{secdistmeasures}. Section \ref{secadscft} describes the applications of this formalism to AdS/CFT. Apart from the application of our measures to coarse-grained subregion dualities, we also discuss fine-grained dualities and the extent to which they can be derived in a theory of quantum gravity. Section \ref{secadscft} is somewhat independent from the rest of the paper. The reader who is interested in our formalism only from an information-theoretic perspective may skip this section. Conversely, this section can also be read independently of the information-theoretic content in this paper for some simple new results in bulk reconstruction. In Appendix  \ref{appsample}, we provide some sample calculations to illustrate our formalism. 

\paragraph{\bf Relation to previous work \\}
We conclude this introduction by reminding the reader of some previous work in these directions. In the holographic context, \cite{Balasubramanian:2013lsa} first discussed the question of evaluating the entropy relevant for an observer who had access to information only in a bounded spacetime region. The question of reconstructing the bulk given only restricted data on the boundary was also considered in \cite{Swingle:2014nla,Kelly:2013aja}. These latter papers are also closely related to the seminal work on S-maximization by Jaynes \cite{jaynes1957information,jaynes1957informationII}. 

However, the perspective in this extant literature differs somewhat from ours. In particular, the principle of S-maximization suggests that given a set of coarse-grained observables, we should evaluate quantum information measures by finding the {\em fine-grained} density matrix with the largest von Neumann entropy that is also consistent with the expectation values of these coarse-grained observables. However, this leads to answers that are sensitive to the structure of the full fine-grained Hilbert space of the theory including, especially, its dimension. This is the reason that the holographic prescriptions above lead to entropies that scale with the central charge of the boundary theory.

In this paper, we would like to consider an observer who has no prior information about the fine-grained degrees of freedom in the theory. This is why the information measures that we define below, in a holographic context, will not scale with the central charge of the boundary theory but just depend on the dimension of the set of restricted observables that we choose to include. 
 
\section{Setup \label{secsetup}}

In this section, we describe our physical setup.  We consider a general physical system in a state $\st$. We also consider an observer, who has
limited abilities to manipulate the system and make measurements. First, we allow the observer to access any linear combination of a set of simple operators  $\al_1 \ldots \al_{\dima}$. This set of operators, which includes the identity operator, is denoted by
\be
\label{alsetdef}
\alset = \text{span~of~}\{\al_1, \ldots \al_{\dima} \}.
\ee  
The operators $\al_i$ need not be Hermitian but we will demand that $\alset$ is closed under the adjoint operation
\be
\al \in \alset \implies \al^{\dagger} \in \alset.
\ee
However, note that when we make reference to a basis for $\alset$ as  in \eqref{alsetdef},  the Hermitian conjugates of all operators are assumed to be included in this basis and we do not need to display them separately.

Note that $\alset$ is {\em not} a $C^{*}$-algebra because\footnote{A simple point, which should not cause confusion is that the set of {\em observables} in $\alset$ consists of its self-adjoint elements. Even usually, the set of observables is not closed under products because the product of two self-adjoint operators may not be self-adjoint. Here, the non-trivial feature is that even the complex vector space spanned by the observables is not closed under products.}
\be
(\al_i \in \alset) \wedge  (\al_j \in \alset)  \centernot \implies \al_i \al_j \in \alset.
\ee

In this paper, we will assume that the complete set of physical quantities that are accessible to the observer is spanned by the {\em two-point} functions
\be
\label{twopointfn}
g_{i j} = \st(\al_i^{\dagger} \al_j).
\ee
We now describe some simple properties of this set of two-point functions.

First, note that since $1 \in \alset$ this set of two-point functions also includes all one-point functions $\st(\al) $. We also assume that the state is normalized so that $\st(1) = 1$. In general, we will assume that that the state $\st$ is {\em separating} with respect to $\alset$ which can be stated as
\be
\label{stseparating}
\st(\al^{\dagger} \al) > 0, \quad  \forall \al \in \alset.
\ee
At times we may consider states where $\st(\al^{\dagger} \al) = 0$ for some $\al \in \alset$ (see, for example, the discussion of pure states below) but these cases can be handled by taking a limit of states that satisfy \eqref{stseparating}.

Second, note that $\alset$ itself is a complex vector space, and when we wish to consider operations on this vector space, we will denote  elements of $\alset$ using bra-ket notation as $|\al \rangle$. This set of states includes $|\onest \rangle$.  The two-point functions \eqref{twopointfn} establish an inner-product on this vector space.
\be
\label{hstinner}
\langle \al_i | \al_j \rangle  = (|\al_i \rangle, |\al_j \rangle) = \st(\al_i^{\dagger} \al_j) = g_{i j}.
\ee
Note that \eqref{stseparating} automatically implies that $g_{i j}$ is Hermitian. To see this, we note that both $\st((\al_i + i \al_j)^{\dagger}(\al_i + i \al_j))$ and $\st((\al_i + \al_j)^{\dagger} (\al_i + \al_j))$ must be real and this can only happen if
\be
g_{j i} = \st(\al_j^{\dagger} \al_i) = \left[\st(\al_i^{\dagger} \al_j)\right]^* = g_{i j}^*.
\ee
Therefore, the inner-product \eqref{hstinner} is conjugate bilinear and positive-definite and endows $\hst$ with the structure of a Hilbert space. We will call $\hst$ the {\em little Hilbert space}.

The ``little Hilbert space''  was first introduced in \cite{Papadodimas:2013jku, Papadodimas:2013wnh}. Here, this space was interpreted as the space of ``small deformations'' about the state $\st$ produced by ``acting'' with elements of $\alset$. The two-point functions \eqref{twopointfn} then just tell us about the ``angles'' between these deformations. In \cite{Almheiri:2014lwa}, the little Hilbert space was termed the ``code subspace''. We will not need these interpretations in this analysis but they are useful to keep in mind.

Note that there is no natural basis choice for $\alset$. We can always make $GL({\dima}, {\cal C})$ transformations on the basis to obtain a new metric. Under such a transformation, the metric transforms as follows
\be
\label{gldctransform}
|\al'_i \rangle = \sum_j M_{q i} |\al_q \rangle \implies g'_{i j} = \sum_{q, k} (M_{q i}|\al_q \rangle, M_{k j} |\al_k \rangle)  = \sum_{q, k} M^{\dagger}_{i q} g_{q k}  M_{k j}.
\ee 
We can use this freedom to always diagonalize $g$. This corresponds to choosing a basis for $\alset$ that satisfies $g_{i j} =  \lambda_i \delta_{i j}$, for some set of real numbers $\lambda_i > 0$. Note that the $\lambda_i$ can themselves be changed by rescaling the basis elements. We will see later how it is possible to extract invariants by combining the adjoint map and the metric.

The description above completely outlines the mathematical framework that we will need in the rest of the paper. However, to remove any possible confusion, we give a simple example.  Consider a lattice quantum field theory, comprising a single local scalar field $\phi(t, x_i)$  on a lattice where the spatial coordinates can take on $N$ distinct values $x_1 \ldots x_N$. Then one interesting restricted set of observables is obtained by taking $\alset$ to be the set of polynomials with products of at most $M$ distinct field operators, with $M \ll N$, on a constant time surface, $t = 0$.  A basis for $\alset$ is given by  monomials of the form
\be
\phi(0,x_{\pi_1}) \ldots \phi(0,x_{\pi_q}),
\ee
where $x_{\pi_1} \ldots x_{\pi_q}$ is any selection of q points of the $N$ available lattice points and $q \leq M$.  Here, we clearly have ${\dima} = \sum_{q = 0}^M {N! \over (N-q)! q!}$. Note that one- and two-point functions of elements of $\alset$ are {\em not} one- and two-point functions in terms of the elementary field; this set contains up to $2 M$-point correlators of the fields.  Second note that $\alset$ is not an algebra because if we multiply two $M$-point monomials, we get a monomial with $2 M$ insertions of the field, and this is not part of $\alset$.

This kind of setup is appropriate for a  bulk observer in AdS/CFT. This observer can measure correlators with a small number of insertions of local bulk fields. However, in general, it is not possible for the observer to measure arbitrarily complicated polynomials of these local bulk fields or measure fields at arbitrarily small separations. The bulk observer also has limited knowledge about the structure of the CFT Hilbert space, and can only operate in the little Hilbert space that she can investigate by exciting the state with sources dual to simple bulk operators.

\subsection{Subtleties with a density matrix interpretation}
When $\alset$ is an algebra, we can identify the density matrix of the system as an element of the algebra, $\rho \in \alset$, with the property that $\tr(\rho \al) = \st(\al), \forall \al \in \alset$.  

However, in the setup under consideration, where we have access only to the little Hilbert space, this is not a viable option. For example, insisting that the density matrix gives the right two-point functions would lead us to demand $\tr(\rho \al_i^{\dagger} \al_j) = g_{i j}$. First, to evaluate the contribution to the trace from within the little Hilbert space, we need to know {\em three-point functions} of elements of $\alset$, which we do not have access to. But even worse since, in general $\al_i \al_j \notin \alset$, the trace also receives contributions from {\em outside} the little Hilbert space. So to evaluate the trace we need to know the action of elements of $\alset$  on the full Hilbert space and not just in the neighbourhood of the state $\st$. This would violate our physical presumption that the observer does not have information about the global Hilbert space or the behaviour of $\al_i$ outside of a neighbourhood of $\st$.

Therefore, here, we will not attempt to construct a density matrix to reproduce the expectation values \eqref{twopointfn}.

\subsection{Pure and mixed states}
We can now proceed to define the notion of a pure state.  A state $\st$ is said to be {\em pure} on the observables $\alset$ if $\st$ has the property that there is no other state $\st'$ that is uniformly smaller than $\st$. We denote the set of pure states by $\Spure$.
\be
\label{purestatedef}
\st \in \Spure \iff \nexists \st' \neq \st, \lambda > 0 ,~\text{such~that}~ \lambda \st'(\al^{\dagger} \al) \leq \st (\al^{\dagger} \al), \forall \al \in \alset.
\ee

This can equivalently be stated as the criterion that a state $\st$ is said to be pure on the observables $\alset$ if it cannot be written as a convex combination of two other states\footnote{When we take linear combinations of states, we mean $(\lambda \st' + (1 - \lambda) \st'')(\al_i \al_j) \equiv \lambda \st'(\al_i \al_j) + (1 - \lambda) \st''(\al_i \al_j)$. Note there are {\em no} cross-terms.}
\be
\st \in \Spure \iff   \st \neq \lambda \st' + (1 - \lambda) \st'',
\ee
with $0 < \lambda < 1$ and any distinct states, $\st', \st'' \neq \st$. 

Note that all the states that appear in these definitions are assumed to be normalized so that $\st(1) = 1$. These definitions are direct generalizations of the definitions used for pure states in quantum information theory when the set of observables forms an algebra. (See, for instance, definition 5.3.5 in \cite{benatti2009dynamics}.) The task here, as in the more complicated examples we encounter later, is to find the right definition --- from among the many equivalent definitions, which hold when $\alset$ is an algebra --- that can generalize to the case where $\alset$ is not an algebra.   

We remind the reader, who may be unfamiliar with the definitions above, that when $\alset$ is an algebra and we can associate a density matrix with the state then, in some basis, the density matrix of a pure state can be written as $\text{diag}\{1,0,0, \ldots\}$. This makes it clear that one cannot find another density matrix that is uniformly smaller than a pure-state density matrix, and also that a pure-state density matrix  cannot be written as a convex combination of other density matrices.

A simple example of a pure state is as follows. Consider the case where  $\alset = \{1, \al \}$, with $\al^{\dagger} = \al$  and consider a state that satisfies
\be
\st(1) = 1; ~~ \st(\al) = \mu; ~~\st(\al^2) = \mu^2,
\ee
for any real value of $\mu$. We might try and construct a ``smaller'' state through $\lambda \st'(1) = \lambda;  \lambda \st'(\al) = \mu'; \lambda \st'(\al^2) = \kappa$, with $\lambda < 1; \kappa < \mu^2$. But a little thought shows that this does not work. First note that since $\lambda \st'((1 - {\lambda \al \over \mu'})^2) \geq 0$, we must have $\kappa \geq {\mu'^2 \over \lambda}$. Then we find that  $\st((1 - {\al \over \mu})^2) = 0$ but $\lambda \st'((1- {\al \over \mu})^2) = \lambda  - 2 {\mu' \over \mu} + {\kappa \over \mu^2} \geq \lambda (1 - {\mu' \over \lambda \mu})^2 \geq 0$. Moreover, the final inequality is saturated only if $\mu' = \lambda \mu$ and $\kappa' = \lambda \mu^2$ but then we would have $\st' = \st$ which is not allowed. Therefore, we {\em cannot} have $\lambda \psi' < \psi$ for any choice of $\lambda$ and $\st'$.

The set of mixed states is the complement of the set of pure states. It comprises all those sets that can be written as convex combinations of other states. 

\subsection{Separable and entangled states  \label{secprodsep}}
It may happen at times, that the space of observables, $\alset$ admits a direct product factorization so that
\be
\label{dirprodobs}
\alset = \alset_1 \otimes \alset_2, \quad [\al_1, \al_2] = 0, ~~\forall \al_1 \in \alset_1, \al_2 \in \alset_2.
\ee
For example, we may consider two separated systems and then consider separate observations on these systems. Even in gravity, we may consider two different observers who are localized in spacelike separated regions of the asymptotic spacetime. Observations made by these observers commute and if we consider the full set of possible observations this forms the direct-product space above. Of course, we do not assume that either $\alset_1$ or $\alset_2$ are algebras. 

A product state on $\alset$ is a state where all expectation values can be decomposed as a product of expectation values in a state on $\alset_1$ and expectation values in a state on $\alset_2$.
\be
\label{prodstatedef}
\st = \st_1 \otimes \st_2 \iff \st((\al_{1i} \al_{2j}) (\al_{1l} \al_{2m})) = \st_1(\al_{1i} \al_{1l}) \st_2(\al_{2 j} \al_{2 m}),
\ee
with $\al_{1i}, \al_{1l} \in \alset_1$ and $\al_{2 j}, \al_{2 m} \in \alset_2$.

 A convex combination of product states is called {\em separable}. We denote the set of separable states by $\Sep$.
\be
\label{sepstatedef}
\st~\in \Sep \quad \iff \quad \st = \sum_i \lambda_i \st_{1 i} \otimes \st_{2 i},
\ee
with $\lambda_i > 0$ and where the subscript $i$ indicates that the sum can run over arbitrary product states. The provision that $\lambda_i > 0$ is important, since otherwise all states on $\alset$ can be written as linear combinations of product states. A  separable state can be interpreted as a classical mixture of product states. 

A state is said to be {\em entangled}  if it is not separable. In section \ref{secdistmeasures}, we will quantify entanglement. In appendix \ref{appsepent}, we give some examples of separable and entangled  states. In general, the question of determining whether a state is separable or entangled is a difficult one, and in fact, it is known to be a NP-hard problem  \cite{ioannou2007computational}. 

We have considered bipartite splittings of the set of observables here, but the same formalism extends in a natural way to {\em multipartite} splittings of the available observables.

\subsubsection{\bf Local operations and classical communication \label{seclocc}} For later use, we consider an important possible transformation of the state defined by $\st \rightarrow \stlocc$, where
\be
\label{loccdef}
\stlocc(\al_{1 i}^{\dagger} \al_{2 k}^{\dagger} \al_{1 j} \al_{2 l}) = \sum_{n=1}^{\loccn} \sum_{\tilde{i}, \tilde{j}, \tilde{k}, \tilde{l}} [M^{(n)}]^{\dagger}_{\tilde{i} i} M^{(n)}_{j \tilde{j}} [N^{(n)}]^{\dagger}_{\tilde{k} k} N^{(n)}_{l \tilde{l}} \st(\al_{1 \tilde{i}}^{\dagger} \al_{2 \tilde{k}}^{\dagger} \al_{1 \tilde{j}} \al_{2 \tilde{l}}).
\ee
Here $M^{(n)}$ and $N^{(n)}$ are arbitrary invertible matrices indexed by $n$ and the sum over $n$ can run to an arbitrary number, $\loccn$. These matrices must satisfy a completeness condition summarized by $\stlocc(1) = 1$. 
Such a transformation is called a LOCC-transformation in the quantum information literature since it can be achieved by an observer who can access observables in $\alset_1$ and another observer who can access observables in $\alset_2$ making local transformations on the state. However, these local transformations are {\em correlated} between the two observers, and these correlations can be achieved by purely classical communication.

It is clear that LOCC operations map separable states to separable states: $\st \in \Sep \implies \stlocc \in \Sep$. Moreover, by enhancing the set of observables to $\blset = \alset_1 \otimes \alset_2 \otimes \alset_{\text{anc}}$ by using an ancillary set of observables, we can realize the LOCC transformation in terms of a  single {\em transformation} of the form \eqref{gldctransform} on an initial product state $\st \otimes \st_{\text{anc}}$. To see this we choose  $\text{dim}(\alset_{\text{anc}}) = \loccn$ and choose the initial state of the ancillary system to satisfy $\st_{\text{anc}}(\al_{\text{anc},i} \al_{\text{anc}, j}) = \delta_{i j}$.  We then consider the following linear transformation acting on elements of $\blset$
\be
\loccstate(\al_{1 i} \al_{2 k} 1_{\text{anc}})  =  \sum_{n=1}^{\loccn} \sum_{\tilde{i}, \tilde{k}} M^{(n)}_{i \tilde{i}} N^{(n)}_{k \tilde{k}} \al_{1 \tilde{i}} \al_{2 \tilde{k}} \al_{\text{anc}, n} ,
\ee
where $1_{\text{anc}}$ denotes the identity operator in the ancillary system. (This is just the analogue of lifting an operator-sum representation of a superoperator to a unitary transformation in a larger space. See, for example, section 3.2 in \cite{preskill1998lecture}.) Although the equation above does {\em not} completely specify $\loccstate$, since we have not specified how it acts on those  elements of  $\blset$ that involve non-trivial insertions of operators from $\alset_{\text{anc}}$,  we will not need this information.

We now consider the state whose two-point functions are given by
\be
L(\st \otimes \st_{\text{anc}})(\bl_i \bl_j) \equiv (\st \otimes \st_{\text{anc}})(L(\bl_i) L(\bl_j)),
\ee
where $\bl_i \in \blset$ and where we have abused notation slightly by allowing $L$ to act both on the state and on $\blset$.

Then it is clear that when restricted only to elements of $\alset_1 \otimes \alset_2$ this state coincides with the original LOCC-transformed state
\be
\label{locclargerstate}
\stlocc(\al_{1 i}^{\dagger} \al_{2 k}^{\dagger} \al_{1 j} \al_{2 l}) = L(\st \otimes \st_{\text{anc}})(\al_{1 i}^{\dagger} \al_{2 k}^{\dagger} \al_{1 j} \al_{2 l}).
\ee
Therefore, the LOCC operation is just an active version of \eqref{gldctransform} in the space with the ancillary observables included.

\section{The Modular and Relative Modular Operators \label{secmodrelmod}}
We now introduce the modular and relative modular operators. Although these operators can be defined independently, they naturally appear
in the Tomita-Takesaki theory of modular automorphisms of von Neumann algebras \cite{takesaki2006tomita,bropalg}, and our discussion follows this path. Several of the properties of the modular and relative modular operators that we describe in this section are well known in the literature on the Tomita-Takesaki theory, and follow from a simple application of the definitions of these operators. So our objective in this section is only to emphasize that these results continue to be true even when $\alset$ is not an algebra. Moreover, this section serves to collect, in one place, all the properties of these operators that we will need.

\subsection{The modular operator}
First, we introduce the anti-linear map, $\S{\st}: \hst \rightarrow \hst$ that acts on the little Hilbert space through
\be
\label{sdef}
\S{\st} | \al \rangle = | \al^{\dagger} \rangle.
\ee
We can also define the adjoint of this map as usual. This is defined by using the usual rules for obtaining the Hermitian adjoint of an anti-linear operator.
\be
\label{sdaggerdef}
\langle \al_i | \S{\st}^{\dagger} | \al_j \rangle = (|\al_i \rangle, \S{\st}^{\dagger} | \al_j \rangle) = (|\al_j \rangle, \S{\st} |\al_i \rangle ) = (|\al_j \rangle, |\al_i^{\dagger} \rangle) = \langle \al_j | \al_i^{\dagger} \rangle = \st(\al_j^{\dagger} \al_i^{\dagger}).
\ee
The modular operator is then defined through
\be
\label{sdaggers}
\modop{\st} = \S{\st}^{\dagger} \S{\st}.
\ee
The relations above allow us to work out the matrix elements of the modular operator.
\be
\label{deltamatrix}
(\modop{\st})_{i j} \equiv \langle \al_i | \modop{\st} | \al_j \rangle = \langle \al_i | \S{\st}^{\dagger} \S{\st} | \al_j \rangle = \langle \al_i | \S{\st}^{\dagger} | \al_j^{\dagger} \rangle 
= \langle \al_j^{\dagger}| \al_i^{\dagger} \rangle = \st(\al_j \al_i^{\dagger}).
\ee
Note that the formula \eqref{deltamatrix} holds in an {\em arbitrary basis}.
It is clear from \eqref{sdaggers} that the modular operator is positive and Hermitian. We can also see this directly. Positivity follows because
\be
\langle \al | \modop{\st} | \al \rangle = \langle \al^{\dagger} | \al^{\dagger} \rangle = \st(\al \al^{\dagger}) \geq 0.
\ee
Hermiticity follows because
\be
\langle \al_j | \modop{\st} | \al_i \rangle = \langle \al_i^{\dagger} | \al_j^{\dagger} \rangle = (\langle \al_j^{\dagger} | \al_i^{\dagger} \rangle)^* = \langle \al_i | \modop{\st} | \al_j \rangle^*.
\ee

We can write the matrix elements of $\modop{\st}$ more explicitly in terms of the operator $\S{\st}$ and the metric $g_{i j}$.  Even though $\S{\st}$ is anti-linear, in a given {\em basis}, we may write
\be
\S{\st} |\al_j \rangle = |\al_j^{\dagger} \rangle = \sum_i s_{i j} |\al_i \rangle.
\ee
We then have
\be
(\modop{\st})_{i j} = \langle \al_j^{\dagger} | \al_i^{\dagger} \rangle = \sum_{q, k} s_{k i} s_{q j}^* \langle \al_q |\al_k \rangle =  \sum_{q, k} s_{q j}^* g_{q k} s_{k i}.
\ee
\subsubsection{Interpretation of the modular operator when $\alset$ is an algebra \label{secmodalg}}
We now pause in our discussion of the properties of the modular operator to discuss its interpretation in a simple setting
We consider the following 
\begin{assumption}
{\em We take the the Hilbert space to have a bipartite factorization, $\ha \otimes \overline{\ha}$, with $\dim(\ha) = \dim(\overline{\ha})$. We take $\alset$ to be the set of all operators that may act non-trivially on $\ha$ but act  trivially on $\overline{\ha}$. Any state, $\st$,  on $\ha$, can be represented as a pure state on $\ha \otimes \overline{\ha}$. We will denote this pure state by $|\pst \rangle$. The little Hilbert space is just generated by the action of $\alset$ on this state: $\hst = \alset |\pst \rangle$. }
\end{assumption}
In the text below, we will return to this simplified setting, as a means of gaining insights on the structures that we define. 

We then claim that the matrix elements \eqref{deltamatrix} are correct if 
\be
\label{formmodular}
\modop{\st} = \rho \otimes \bar{\rho}^{-1}, \quad \text{when $\alset$ is an algebra as in \specialcase}. 
\ee
This means that $\modop{\st}$ acts as the density matrix on $\ha$ and as the inverse of the density matrix on $\overline{\ha}$. We remind the reader of condition \eqref{stseparating}, which ensures that $\rho$ is nonsingular. 

In terms of the modular Hamiltonian $H^{\text{mod}}_A$ on $\ha$, this means that we can write $\modop{\st} = e^{-H^{\text{mod}}_A + \overline{H}^{\text{mod}}_{A}}$ where $\overline{H}^{\text{mod}}_{A}$ is the modular Hamiltonian for $\overline{\ha}$.  Note that, as a consequence of this relation,  $\modop{\st}$ is not an element of the algebra, and nor is it an element of the commutant, which consists of operators that act trivially on $\ha$. 

We now verify the formula \eqref{formmodular}.  In the {\em Schmidt basis} the state can be written as
\be
|\pst \rangle = \sum a_n |n, \bar{n} \rangle,
\ee
where $|n \rangle$ runs over some orthonormal basis and $|\bar{n} \rangle$ denotes the vector in $\overline{\ha}$ that is entangled with $|n \rangle$.
Then we have
\be
\rho = \tr_{\overline{\ha}} (|\pst \rangle \langle \pst | ) = \sum_{n} |a_n|^2 |n \rangle \langle n|.
\ee
And the claimed modular operator is given by
\be
\modop{\st} = \rho \otimes \bar{\rho}^{-1}  = \sum_{n, \bar{m}} {|a_n|^2 \over |a_m|^2} |n, \bar{m} \rangle \langle n, \bar{m} |. 
\ee
Let us now take two operator $A_p, A_q: \ha \rightarrow \ha$ and lift them to operators on the full space by demanding that they  act as the identity on $\overline{\ha}$. We now see that
\be
\langle \pst | (A_q \otimes \idmat) \modop{\st} (A_p \otimes \idmat)  | \pst \rangle = \sum_{n,r,s,n'} (a_n)^*  {|a_{s}|^2 \over |a_{r}|^2}  a_{n'} \langle n, \bar{n} | A_q \otimes \idmat  |s, \bar{r} \rangle \langle s, \bar{r} | A_p \otimes \idmat | n', \bar{n}' \rangle.
\ee
The dot-product of the barred-vectors just yields the delta functions $\delta_{n r} \delta{r n'}$ leading to 
\be
\langle \pst | (A_q \otimes \idmat) \modop{\st} (A_p \otimes \idmat) | \pst \rangle = \sum_{n, s} |a_s|^2 \langle n | A_q | s \rangle \langle  s | A_p | n \rangle = \langle \pst | (A_p \otimes \idmat)  (A_q \otimes \idmat) | \pst \rangle,
\ee
in precise agreement with \eqref{deltamatrix}.

We also note that the, due to \eqref{stseparating}, the set of states $\alset | \pst \rangle$ is dense in the full Hilbert space, $\ha \otimes \hab$. Therefore the matrix elements of the modular operator above completely specify the operator in $\ha \otimes \hab$.

The result above can also be interpreted as follows. Since the space $\hst$ in this case is isomorphic to $\ha \otimes \hab$ we can also write the action of the modular operator as
\be
\modop{\st} |\al \rangle = |\rho \al \rho^{-1} \rangle, ~\text{when~$\alset$~is~an~algebra}.
\ee
It is easy to see then that 
\be
\langle \al_i | \modop{\st} | \al_j \rangle = \tr(\rho \al_i^{\dagger} \rho \al_j \rho^{-1}) = \tr\big(\rho \al_j \al_i^{\dagger}\big) = \langle \al_j^{\dagger} | \al_i^{\dagger} \rangle,
\ee
precisely as required. Indeed, the modular operator is often introduced as an operator that acts on the space of operators in the Hilbert space. (See, for instance, \cite{nielsen2004simple}.) However, we find that \eqref{formmodular} (which holds under the additional conditions outlined in \specialcase) is more useful in providing intuition about $\modop{\psi}$.

\subsubsection{Spectrum of the modular operator}
The matrix elements of \eqref{deltamatrix} transform under $GL({\dima}, C)$ transformations of the basis that we use for $\alset$. However, we now show how the {\em spectrum} of eigenvalues of $\modop{\st}$ provides us with an invariant quantity that characterizes the state. The spectrum is defined by the usual eigenvalue equation
\be
\label{eigenvaldelta}
  \sum_i c_i \modop{\st} |\al_i \rangle  = \lambda  \sum_i c_i | \al_i \rangle,
\ee
which yields the eigenvectors $\sum_i c_i |\al_i \rangle$ and the eigenvalues $\lambda$.

Note that the matrix elements of the modular operator are given by \eqref{deltamatrix}, but such a basis is not orthonormal in general, so we cannot simply diagonalize \eqref{deltamatrix}. Instead we go to an orthonormal basis denoted by
\be
\label{alodef}
|\al_i^O \rangle = \sum_j M^O_{j i} | \al_j \rangle,
\ee
which satisfies $\langle \al_j^O | \al_i^O \rangle = \delta_{i j}$. Of course the transformation $M^{O}_{j i}$ is not unique since we can make unitary transformations. Any other basis that is related by $|\al_i^{O'} \rangle = \sum_j U_{j i} |\al_j^{O} \rangle$, where $U_{i j}$ is a unitary matrix, also satisfies $\langle \al_i^{O'} | \al_j^{O'} \rangle = \delta_{i j}$.

In the orthonormal basis, the matrix elements of the modular operator are given by
\be
(\modop{\st})^O_{i j} = \langle \al_i^{O} | \modop{\st} | \al_j^{O} \rangle = \sum_{q, l}  (M^{O})^{\dagger}_{i q} \modop{\st}_{q l} M^{O}_{l j}.
\ee
The eigenvalues of the matrix $(\modop{\st})^O_{i j}$ give us the spectrum that we need. It is easy to see that this spectrum is invariant under the unitary ambiguity that exists in \eqref{alodef}. In the primed basis above, we would have
\be
(\modop{\st})^{O'}_{i j}  = \sum_{k, l} U^{\dagger}_{k i} U_{l j} (\modop{\st})^O_{k l},
\ee
and this matrix clearly has the same spectrum as $\modop{\st}^{O}$. 

We can also solve \eqref{eigenvaldelta} directly by simply contracting it with  $\langle \al_j | $ leading to the equation
\be
\sum_i c_i (\modop{\st})_{j i} = \lambda \sum_i c_i g_{j i}.
\ee 
If we denote the inverse of $g$ by $g^{-1}$ so that $\sum_j g^{-1}_{k j} g_{j i} = \delta_{k i}$ then the equation above becomes
\be
\sum_{i,j} c_i g^{-1}_{k j} (\modop{\st})_{j i} = \lambda c_k.
\ee
We can now diagonalize the matrix  $\sum_j g^{-1}_{k j} \modop{\st}_{j i}$ to obtain the spectrum of eigenvalues $\lambda$. Under a $GL({\dima}, C)$ transformation as in \eqref{gldctransform} we see that
\be
(g')^{-1}_{i j} =  \sum_{l, k} M^{-1}_{i l} g^{-1}_{l k} (M^{\dagger})^{-1}_{k j}; \quad (\modop{\st}')_{i j} = \sum_{l, k} M^{\dagger}_{i l} \modop{\st}_{l k} M_{k j} ,
\ee
and
\be
\sum_{i} (g')^{-1}_{k i} (\modop{\st}')_{i j} = \sum_{l, i, q} (M)^{-1}_{k l} g^{-1}_{l i} \modop{\st}_{i q} M_{q j}.
\ee
Since matrices related by similarity transformations have the same spectrum, it is clear that the two matrices above have the same spectrum. This argument also shows that the spectrum of eigenvalues thus obtained is the same as the spectrum of $(\modop{\st})^{O}_{i j}$ introduced above. 

Therefore the spectrum of the modular operator characterizes the state independent of the basis chosen for $\alset$.

\subsubsection{Pairing of eigenvalues of the modular operator \label{secmodpairing}}
We now show that eigenvalues in the spectrum of the modular operator appear in reciprocal pairs. This can be written as
\be
\spect\big(\modop{\st}\big) = \spect{\big((\modop{\st})^{-1}\big)},
\ee
where we use the notation $\spect$ to denote the spectrum of an operator.

We start by proving the identity
\be
\label{deltathroughs}
\S{\st} \modop{\st} = (\modop{\st})^{-1} \S{\st}.
\ee
Note that from \eqref{sdef} we obviously have $\S{\st}^2 = 1$.  We also have $(\S{\st}^{\dagger})^2 = 1$ since for arbitrary $\al_i$ and $\al_j$,
\be
 (|\al_j \rangle, \S{\st}^{\dagger} \S{\st}^{\dagger} |\al_i \rangle ) = (\S{\st} |\al_j \rangle, \S{\st}^{\dagger} |\al_i \rangle)^* = (\S{\st}^2 |\al_j \rangle, |\al_i \rangle) = (|\al_j \rangle, |\al_i \rangle).
\ee
Now multiplying both sides of \eqref{deltathroughs} by $\S{\st} \modop{\st}$ we see that the relation we need to prove is
\be
\S{\st} \modop{\st} \S{\st} \modop{\st} = 1.
\ee
But using the definition of $\modop{\st}$ we see that this is just
\be
\S{\st} \modop{\st} \S{\st} \modop{\st}  = \S{\st} (\S{\st}^{\dagger} \S{\st}) \S{\st} (\S{\st}^{\dagger} \S{\st}) = 1.
\ee

Now consider a solution to the eigenvalue equation for $\modop{\st}$ \eqref{eigenvaldelta}
\be
\modop{\st} |\al \rangle = \lambda |\al \rangle.
\ee
Then we have
\be
\modop{\st} |\al^{\dagger} \rangle = \modop{\st} \S{\st} |\al \rangle = \S{\st} (\modop{\st})^{-1} |\al \rangle = \S{\st} \lambda^{-1} |\al \rangle = \lambda^{-1} |\al^{\dagger} \rangle,
\ee
where, in the last equation, we have used the fact that all eigenvalues of $\modop{\st}$ are real and positive. Therefore $\S{\st} |\al \rangle$ is an eigenvector of $\modop{\st}$ with eigenvalue $\lambda^{-1}$.

Note also that if $|\al \rangle$ is an eigenvector of $\modop{\st}$ and if $\al^{\dagger} = \al$ then we must have $\lambda = 1$.

The fact that eigenvalues of $\modop{\st}$ appear in reciprocal pairs is natural in the situation where $\alset$ is an algebra acting on one part of a bipartite Hilbert space as in \specialcase. We see from \eqref{formmodular} that if $\rho_i$ are the eigenvalues of $\rho$ then the eigenvalues of $\modop{\st}$ are just ${\rho_i \over \rho_j}$ and so they naturally occur in reciprocal pairs corresponding to all possible ratios of the eigenvalues of the density matrix. But we see above that even when $\alset$ is not an algebra, the modular operator continues to have this property.

\subsection{The relative modular operator}
The modular operator characterizes a state. We now describe the {\em relative modular operator} which depends on two states and can be used to characterize their difference. To define the relative modular operator, we consider a {\em second} state $\sst$. This second state also induces a separate Hilbert space structure on $\alset$ leading to another little Hilbert space $\hsst$. We denote vectors in this space through $|\al \rangle_{\sst}$ and the inner-product is given by
\be
 _{\sst} \langle \al_i | \al_j \rangle_{\sst} = \sst(\al_i^{\dagger} \al_j).
\ee
Note that we denote vectors in $\hsst$ with an additional subscript. But, in this section,  we will continue to denote vectors in $\hst$ using the notation $|\al \rangle$ with no additional subscripts.

Then we define a map $\rels{\st}{\sst}: \hst \rightarrow \hsst$ through
\be
\rels{\st}{\sst} |\al \rangle = |\al^{\dagger} \rangle_{\sst}.
\ee
The adjoint of this map $\relsd{\st}{\sst}: \hsst \rightarrow \hst$ is defined through
\be
\langle \al_i | \relsd{\st}{\sst} | \al_j \rangle_{\sst} = {}_{\sst}\langle \al_j | \rels{\st}{\sst} | \al_i \rangle =  {}_{\sst}\langle \al_j | \al_i^{\dagger} \rangle_{\sst} = \sst(\al_j^{\dagger} \al_i^{\dagger}).
\ee
The relative modular operator is then defined through
\be
\label{defrelmods}
\relmod{\st}{\sst} = \relsd{\st}{\sst} \rels{\st}{\sst}.
\ee
From the definitions above, it is clear that $\relmod{\st}{\sst}$ maps $\hst \rightarrow \hst$ and its matrix elements are
\be
\langle \al_j | \relmod{\st}{\sst} | \al_i \rangle = {}_{\sst} \langle \al_i^{\dagger} | \al_j^{\dagger} \rangle_{\sst} = \sst(\al_i \al_j^{\dagger}) .
\ee

\subsubsection{Interpretation of the relative modular operator when $\alset$ is an algebra \label{secrelmodalg}}
We now describe the interpretation of the relative modular operator in the special case \specialcase. We additionally assume that the state $\sst$ is also pure in the full Hilbert space and we denote this pure state by $|\psst \rangle$.  As usual,  we denote the density matrix of $\ha$ in the state $\st$ by $\rho$  and remind the reader that the density matrix of $\overline{\ha}$ is then $\overline{\rho}$ with the same spectrum as $\rho$.  Similarly, we denote the density matrix of $\ha$ in the state $\sst$ by $\sigma$.  We will then show that
\be
\label{relmodalg}
\relmod{\st}{\sst} = \sigma \otimes \overline{\rho}^{-1}, \quad \text{~in~the~special~case~\specialcase}.
\ee
Here, as in \eqref{formmodular}, the second factor acts on the complementary space.

We assume that both states have a Schmidt decomposition as follows
\be
\label{densscrelmod}
\begin{split}
&|\pst \rangle = \sum_n a_n |n, \bar{n} \rangle, \\
&|\psst \rangle = \sum_{\alpha} b_{\alpha} |\alpha, \bar{\alpha} \rangle.
\end{split}
\ee
Notice that the Latin indices, $n$ and the Greek indices $\alpha$ run over {\em different} sets of states, corresponding to the Schmidt basis that diagonalizes entanglement in the two states. The density matrices are given by
\be
\begin{split}
&\rho  = \sum_n |a_n|^2 |n \rangle \langle n|, \\
&\sigma = \sum_{\alpha} |b_{\alpha}|^2 |\alpha \rangle \langle \alpha|.
\end{split}
\ee
and the claimed form of the relative modular operator is
\be
\relmod{\st}{\sst} = \sum_{\alpha, n} {|b_{\alpha}|^2 \over |a_n|^2} |\alpha, \bar{n} \rangle \langle \bar{n}, \alpha |.
\ee
We see, using the form above, that if $\al_p, \al_q$ are any two operators: $\ha \rightarrow \ha$ then 
\be
\begin{split}
& \langle \pst | (\al_p \otimes \idmat) \relmod{\st}{\sst} (\al_q \otimes \idmat)  | \pst \rangle = \sum_{n,m, n', \alpha} a_n^* a_{n'} {|b_{\alpha}|^2 \over |a_m|^2} \langle n, \bar{n} | \al_p \otimes \idmat  |\alpha, m \rangle \langle \alpha, m | \al_q \otimes \idmat | n', \bar{n}' \rangle \\ &=
\sum_{n, \alpha} \langle n | \al_p | \alpha \rangle |b_{\alpha}|^2 \langle \alpha | \al_ q | n \rangle = \sum_{\alpha} |b_{\alpha}|^2 \langle \alpha | \al_q \al_p | \alpha \rangle = \langle \psst| \al_q \al_p | \psst \rangle,
\end{split}
\ee
which is precisely what we need. Note that in going from the first to the second line, we used the identity operators to equate $n = m = n'$.

Just as in section \ref{secmodalg} the space $\hst$ is isomorphic to $\ha \otimes \hab$ due to the separating nature of the state, $\st$. This has two implications. First, it tells us that the matrix elements above completely specify the relative modular operator in the special case \specialcase. Second, it allows us to rewrite this result  in the slightly more general form
\be
\relmod{\st}{\sst} | \al \rangle = |\sigma \al \rho^{-1} \rangle, ~\text{when~$\alset$~is~an~algebra}.
\ee
We can easily verify that, in this case,
\be
\langle \al_i | \relmod{\st}{\sst} | \al_j \rangle = \tr(\rho \al_i^{\dagger} \sigma \al_j \rho^{-1}) = \tr(\sigma \al_j \al_i^{\dagger}) = \sst(\al_j \al_i^{\dagger}),
\ee
precisely as required. Once again, the reader may find the result \eqref{relmodalg} (which requires additional conditions) somewhat more useful to develop intuition about the relative modular operator.

\paragraph{\bf Spectrum of the relative modular operator \\}
 We define the eigenvalues of the relative modular operator through the eigenvalue equation
\be
\relmod{\st}{\sst} | \al \rangle = \lambda |\al \rangle.
\ee

The spectrum of possible values of $\lambda$ can be computed by going to an orthogonal basis, just as we described for the modular operator. Once again, this spectrum does not depend on the choice of basis for $\alset$. We will not repeat the proof of this claim, since this discussion is almost identical to the discussion for the modular operator above.

\subsubsection{ Relationship between the eigenvalues of $\relmod{\st}{\sst}$ and $\relmod{\sst}{\st}$ \label{secrelmodinv}}
We now show that the spectrum of $\relmod{\st}{\sst}$ is the {\em inverse} of the spectrum of $\relmod{\sst}{\st}$:
\be
\spect\left(\relmod{\st}{\sst}\right) = \spect \left(\relmod{\sst}{\st}^{-1}\right).
\ee
In the special case \specialcase, this property is obvious from \eqref{relmodalg} but we will show that it holds more generally.

To see this we note that the maps $\rels{\st}{\sst}: \hst \rightarrow \hsst$ and $\rels{\sst}{\st}: \hsst \rightarrow \hst$ satisfy
\be
\rels{\sst}{\st} \rels{\st}{\sst} = 1_{\st}, \quad \rels{\st}{\sst} \rels{\sst}{\st} = 1_{\sst},
\ee
where the subscript on $1$ distinguishes the identity operators in the two spaces. It follows that
\be
\label{relsinvrelation}
\relsd{\st}{\sst} \relsd{\sst}{\st} = 1_{\st}, \quad \relsd{\sst}{\st} \relsd{\st}{\sst} = 1_{\sst}.
\ee

Some simple manipulations of these identities leads to 
\be
\begin{split}
1_{\st} &= \rels{\st}{\sst}^{\dagger}  \rels{\sst}{\st}^{\dagger} =  \rels{\st}{\sst}^{\dagger} (\rels{\st}{\sst} \rels{\sst}{\st}) \rels{\sst}{\st}^{\dagger} (\rels{\sst}{\st} \rels{\st}{\sst}) \\
&= (\rels{\st}{\sst}^{\dagger} \rels{\st}{\sst}) \rels{\sst}{\st} (\rels{\sst}{\st}^{\dagger}\rels{\sst}{\st}) \rels{\st}{\sst} = \relmod{\st}{\sst} \rels{\sst}{\st} \relmod{\sst}{\st} \rels{\st}{\sst},
\end{split}
\ee
where, in the last step, we used the definition \eqref{defrelmods}.But this means that
\be
\label{relmodscommut}
\relmod{\st}{\sst} \rels{\sst}{\st} = \rels{\sst}{\st} \relmod{\sst}{\st}^{-1}.
\ee

The relation \eqref{relmodscommut} now tells us that given an eigenvector of $\relmod{\sst}{\st}^{-1}$ that satisfies
\be
\relmod{\sst}{\st}^{-1} |\al \rangle_{\sst} = \lambda | \al \rangle_{\sst},
\ee
we have an eigenvector of $\relmod{\st}{\sst}$ with the same eigenvalue:
\be
\relmod{\st}{\sst} \rels{\sst}{\st} | \al \rangle_{\sst} = \rels{\sst}{\st} (\rels{\st}{\sst} \relmod{\st}{\sst} \rels{\sst}{\st}) | \al \rangle_{\sst} = \lambda \rels{\sst}{\st} |\al \rangle_{\sst},
\ee
where in the last step we have assumed that $\lambda \in {\cal R}^{+}$ and so it commutes with $\rels{\sst}{\st}$.

\subsubsection{Relationship between the relative modular and modular operators \label{secspectrelationship}}
It is clear from their respective definitions that the modular operator and the relative modular operator are related through
\be
\modop{\st} = \relmod{\st}{\st}.
\ee
It is also clear that the state $\sst$ is {\em equal} to the state $\st$  only if $\modop{\st} = \relmod{\st}{\sst}$. 
\be
\label{equalitycondition}
\langle \al_i | \modop{\st} | \al_j \rangle = \langle \al_i| \relmod{\st}{\sst} | \al_j \rangle, \forall \al_i, \al_j \iff \st(\al_i^{\dagger} \al_j) = \sst(\al_i^{\dagger} \al_j),  \forall \al_i, \al_j.
\ee

We will also need one additional property in what follows. The spectrum of $\modop{\st} \relmod{\st}{\sst}^{-1}$ is the same as the spectrum of  $\relmod{\sst}{\st} \modop{\sst}^{-1} $. 
This is natural when we consider the special case \eqref{relmodalg}. However, the result holds more generally and we can demonstrate it as follows.

First we note that 
\be
\S{\st} = \rels{\sst}{\st} \S{\sst} \rels{\st}{\sst},
\ee
as can be easily checked through their action on a general state. Correspondingly, we also have
\be
\S{\st}^{\dagger} = \rels{\st}{\sst}^{\dagger} \S{\sst}^{\dagger} \rels{\sst}{\st}^{\dagger}.
\ee
Therefore
\be
\modop{\st} = \S{\st}^{\dagger} \S{\st} = \rels{\st}{\sst}^{\dagger} \S{\sst}^{\dagger} \rels{\sst}{\st}^{\dagger}\rels{\sst}{\st} \S{\sst} \rels{\st}{\sst} = \rels{\st}{\sst}^{\dagger} \S{\sst}^{\dagger} \relmod{\sst}{\st}  \S{\sst} \rels{\st}{\sst},
\ee
and using the fact that 
\be
\relmod{\st}{\sst}^{-1} = \rels{\sst}{\st} \rels{\sst}{\st}^{\dagger},
\ee
which follows from \eqref{relsinvrelation}, we find that
\be
\begin{split}
\modop{\st} \relmod{\st}{\sst}^{-1} &= \rels{\st}{\sst}^{\dagger} \S{\sst}^{\dagger} \relmod{\sst}{\st}    \S{\sst} \rels{\st}{\sst} \rels{\sst}{\st} \rels{\sst}{\st}^{\dagger} = \rels{\st}{\sst}^{\dagger} \S{\sst}^{\dagger}  \relmod{\sst}{\st} \S{\sst} \rels{\sst}{\st}^{\dagger} \\
&=  \rels{\st}{\sst}^{\dagger} \S{\sst}^{\dagger}  \relmod{\sst}{\st} \S{\sst} \S{\sst}^{\dagger} \S{\sst}^{\dagger} \rels{\sst}{\st}^{\dagger} =  \rels{\st}{\sst}^{\dagger} \S{\sst}^{\dagger}  \relmod{\sst}{\st} \modop{\sst}^{-1} \S{\sst}^{\dagger} \rels{\sst}{\st}^{\dagger}.
\end{split}
\ee
But now we see that 
\be
\begin{split}
&\modop{\st} \relmod{\st}{\sst}^{-1} |\al \rangle = \lambda | \al \rangle \\
\implies  &\relmod{\sst}{\st} \modop{\sst}^{-1} \left(\S{\sst}^{\dagger} \rels{\sst}{\st}^{\dagger} | \al \rangle \right) =  \left(\S{\sst}^{\dagger} \rels{\sst}{\st}^{\dagger}   \rels{\st}{\sst}^{\dagger} \S{\sst}^{\dagger} \right)  \relmod{\sst}{\st} \modop{\sst}^{-1} \left(\S{\sst}^{\dagger} \rels{\sst}{\st}^{\dagger} | \al \rangle \right) \\ = &\lambda \left(\S{\sst}^{\dagger} \rels{\sst}{\st}^{\dagger} | \al \rangle  \right).
\end{split}
\ee
Using the fact that the spectrum of the operator is left unchanged by a similarity transform, this result easily extends into the following result. 
\be
\label{spectrumsame}
\spect \left(
(\modop{\st})^{1-x} (\relmod{\st}{\sst})^{-1} (\modop{\st})^{x} \right) = \spect\left((\modop{\sst})^{-y} \relmod{\sst}{\st} (\modop{\sst})^{y-1}\right),
\ee
for arbitrary values of $x,y$.

\section{Measures of Distance and Entanglement \label{secdistmeasures}}
We have now set the stage for defining quantum information measures for sets.  In quantum information theory, an important role in defining measures of information is played by a notion of distance between states. 

When $\alset$ is an algebra,  the relative entropy is a commonly used notion of distance between states. However, we describe how some simple attempts to directly generalize the notion of relative entropy fail. Then we describe some new measures of distance between states for sets of observables and show that these measures meet our requirements.

We will  look for a measure of distance $\gendist{\alset}{\st}{\sst}$ that depends on the observables at hand --- two-point functions of elements of  $\alset$ --- and satisfies a set of properties. The significance of these properties is that it was shown in \cite{vedral1997quantifying} that a measure of distance that satisfies properties (1)-(3) can be used to define a {\em good measure of bipartite  and multipartite entanglement} in a sense that we will review below. 

We demand the following properties from our distance measure $\gendist{\alset}{\st}{\sst}$.
\begin{enumerate}
\item{\bf Basis Independence \label{propbasisind}}
A very basic property, that we will demand from all distance measures, is that they should not depend on the basis chosen for $\alset$.
\item{\bf Specificity: \label{proppos}}
In general, we would like distance to be positive: $\gendist{\alset}{\st}{\sst} \geq 0$. However, it is important that the measure be {\em specific}, in that
$\gendist{\alset}{\st}{\sst} = 0 \iff \st = \sst$. Here equality between states means that one-point and two-point functions of elements of $\alset$ are the same.
\item{\bf Monotonicity: \label{propmon}}
Reducing the number of observables with which one probes the state should make states less distinguishable. Therefore if $\blset \subset \alset$ then $\gendist{\blset}{\st}{\sst} \leq \gendist{\alset}{\st}{\sst}$. 
\item{\bf Insularity: \label{propins}}
The distance between two states should not change if we simply add on a spectator system. If we write $\alset = \alset_1 \otimes \alset_2$ as in \eqref{dirprodobs} and we consider states $\st=\st_1 \otimes \st_2$ and $\sst = \sst_1 \otimes \st_2$ (where product states are defined in \eqref{prodstatedef}) then
\be
\gendist{\alset}{\st}{\sst} = \gendist{\alset_1}{\st_1}{\sst_1}.
\ee
\end{enumerate}
We call property \eqref{propins} ``insularity'' because for the distance measure to be meaningful it should not ``care'' about the state of the rest of the Universe. From a physical perspective, this is an obvious property that a measure of distance must obey but we will see that it is surprisingly effective in ruling out several plausible distance-measures.

It is often useful to have a measure of distance  that satisfies two additional nice properties: {\em additivity} and {\em finiteness}. We emphasize that these properties are often {\em not necessary}; indeed, even in the algebraic setting, the relative entropy is not finite.
\begin{enumerate}
\setcounter{enumi}{4}
\item{\bf Additivity: \label{propaddit}}
For a product state as defined in \eqref{prodstatedef},  we would like
\be
\gendist{\alset_1 \otimes \alset_2}{\st_1 \otimes \st_2}{\sst_1 \otimes \sst_2} = \gendist{\alset_1}{\st_1}{\sst_1} + \gendist{\alset_2}{\st_2}{\sst_2}.
\ee
This is useful, particularly if we use the distance measure to define a notion of {\em entropy}, which needs to be extensive. 
\item{\bf Finiteness: \label{propfinite}} This is simply the property that the distance between any two states is bounded above. 
\be
\exists K, \text{such~that} \gendist{\alset}{\st}{\sst} < K, \forall \st, \sst.
\ee
\end{enumerate}

\subsection{\bf Distance and Entanglement \label{disteng}}
We now explain the relationship between measures of distance and entanglement following \cite{vedral1997quantifying} . We will need to introduce additional structure on the space of observables to make sense of the notion of entanglement. In this subsection, we assume that the set of accessible observables satisfies a direct-product structure as in \eqref{dirprodobs}. This helps us define a notion of {\em bipartite entanglement}.  We set this measure $\ent{\alset}{\st}$ to be
\be
\label{generalentangdef}
\ent{\alset}{\st} = \underset{\sst \in \Sep}{\text{inf}} \gendist{\alset}{\st}{\sst},
\ee
where $\sst$ runs over the set of {\em separable states} as defined in \eqref{sepstatedef}. A notion of {\em multipartite entanglement} can be defined along similar lines.

Intuitively speaking, this measure is a generalization of the ``mutual information.'' We remind the reader that the mutual information can be defined as the relative entropy between a state and the corresponding product state. More precisely, given a state $\st$, we define an associated product state by $\st_{\text{prod}}\left({\al_{1 i_1} \al_{1 i_2} \al_{2 j_1} \al_{2 j_2}}\right) = \st(\al_{1 i_1} \al_{1 i_2}) \st(\al_{2 j_1} \al_{2 j_2})$. Then, the mutual information is just defined as the relative entropy between these two states: $\relent{\al}{\st}{\st_{\text{prod}}}$. 

So the mutual information measures the distance between a state and a {\em specific} separable state (the associated product state), using a specific notion of distance (the relative entropy). The definition \eqref{generalentangdef} has two differences. First, it does not commit itself to using a particular separable state since it may happen that the product state is {\em not} the closest separable state. Second, it allows us to use an arbitrary notion of distance and this is important for us since, as we explain below, we are unable to satisfactorily generalize the relative entropy when $\alset$ is not  an algebra. 

It was shown in \cite{vedral1997quantifying} that the measure of entanglement defined through \eqref{generalentangdef} satisfies the following desirable properties.
\begin{enumerate}
\item
It measures only {\em quantum correlations}. This is because some correlations between observations in $\alset_1$ and $\alset_2$ can simply be explained through classical probabilistic physics. The states that have only such correlations are precisely the set of {\em separable} states. But if we see that if 
\be
\st \in \Sep \iff \ent{\alset}{\st} = 0.
\ee
 So, $\ent{\alset}{\st}$ gives us a measure of quantum correlations.
\item
$\ent{\alset}{\st}$ is invariant under {\em local} unitary transformations. Here we assume that $\exists U_1 \in \alset_1$ and $\exists U_2 \in \alset_2$ such that $U_1 \al_{1} U_1^{\dagger} \in \alset_1$, $\forall \al_{1} \in \alset_1$  and $U_2 \al_2 U_2^{\dagger} \in \alset_2$, $\forall \al_{2} \in \alset_2$.  Then the observer can {\em act} with $U_1$ and $U_2$ and again make measurements in $\alset_1$ and $\alset_2$.  More precisely,  we change the state to $U_1 U_2(\st)$ which is defined by
\be
U_1 U_2(\st)(\al_{1 i} \al_{1 j}  \al_{2 k}  \al_{2 l}) \equiv \st(U_1 \al_{1 i} \al_{1 j} U_1^{\dagger} U_2 \al_{2 k} \al_{2 l} U_2^{\dagger}) .
\ee
Now we see that if $\sst \in \Sep$ then $U_1 U_2(\sst) \in \Sep$ and moreover that 
\be
\gendist{\alset}{U_1 U_2(\st)}{U_1 U_2(\sst)} = \gendist{\alset}{\st}{\sst},
\ee
which follows from Property \eqref{propbasisind}. So
\be
\ent{\alset}{U_1 U_2(\st)} = \ent{\alset}{\st}.
\ee
and therefore, as expected, entanglement is unchanged by local unitary transformations. 
\item
The entanglement between two subsystems is not changed by simply adding on a separate auxiliary system, and the property \eqref{propins} of the distance ensures this.
\item
Finally, we can consider LOCC operations as discussed in section \ref{seclocc}. We see that entanglement can only {\em decrease} under such operations. This is because we see that in evaluating $\ent{\alset}{\stlocc}$, if the infimum of \eqref{generalentangdef}  is attained at $\sst$ then $\ent{\alset}{\stlocc} \leq \gendist{\alset}{\stlocc}{\sstlocc}$, since $\sstlocc \in \Sep$. But since, as explained in section \ref{seclocc}, the LOCC-operation can be achieved by a simple basis-transformation in an enlarged space with ancillary observables
\be
\gendist{\alset}{\stlocc}{\sstlocc} \leq \gendist{\alset \otimes \alset_{\text{anc}}}{L(\st \otimes \st_{\text{anc}})}{L(\sst \otimes \st_{\text{anc}})} = \gendist{\alset}{\st}{\sst}.
\ee
Here, in the first step we used the monotonicity of the distance function and in the next step we used the basis-independence and insularity of the distance function. Therefore we see that  $\ent{\alset}{\stlocc} \leq \ent{\alset}{\st}$.

This is a pleasing result since it shows that entanglement is fundamentally a quantum resource that can only be destroyed and not created by two observers acting locally and communicating classically.

\end{enumerate}

So we see that an appropriate measure of distance and the notion of separable states is essential to quantifying entanglement.

\subsection{Distance measures}
In the rest of the section, we now discuss measures of distance that satisfy the properties above. 

When $\alset$ is an algebra, the relative entropy is a very useful notion of distance. So,  we will first discuss some attempts to construct analogues of the relative entropy by generalizing a construction due to Araki \cite{araki1976relative}. The measures constructed in this manner reduce to the relative entropy when $\alset$ is an algebra but when $\alset$ is not an algebra they either fail to satisfy property \eqref{proppos} (specificity) or else they fail to satisfy property \eqref{propmon} (monotonicity). 

Therefore, in subsection \ref{secnormentropy}, we then construct new measures of distance that satisfy all the necessary properties \eqref{propbasisind} (basis independence), \eqref{proppos} (specificity), \eqref{propmon} (monotonicity) and  \eqref{propins} (insularity) .  One of these measures --- the ``normed entropy'' --- also satisfies property \eqref{propaddit} (additivity) although it does not satisfy property \eqref{propfinite} (finiteness). We also find a large class of measures that do not satisfy property \eqref{propaddit} but satisfy property \eqref{propfinite}.

\subsubsection{Generalizations of the relative entropy}
When $\alset$ is an algebra, Araki \cite{araki1976relative} showed that the usual relative entropy is just given by
\be
\label{relativearaki}
\relent{\alset}{\st}{\sst} = -\langle \onest |\log\big[\relmod{\st}{\sst}\big] | \onest \rangle.
\ee
It is easy to verify this formula under the assumptions of \specialcase, when $\alset$ is an algebra and when $\st$ and $\sst$ can be represented by density matrices $\rho$ and $\sigma$ respectively. In that situation recall that the standard definition of the relative entropy is 
\be
\label{relativeusual}
\relent{\alset}{\st}{\sst} = \tr\left(\rho  \log(\rho) -  \rho \log(\sigma) \right).
\ee
Now consider the expression for the relative modular operator under the assumptions of \specialcase ~as developed in subsection \ref{secrelmodalg}. From that expression we see that
\be
-\langle \onest | \log\big[\relmod{\st}{\sst}\big] | \onest \rangle = \sum_{n}|a_n|^2 \Big(\log |a_n|^2  - \sum_{\alpha} |\langle \alpha | n \rangle |^2 \log |b_{\alpha}|^2  \Big).
\ee
Using the expressions for the density matrices in \eqref{densscrelmod} it is easy to see that this is precisely the same as \eqref{relativeusual}.

We now consider the quantity defined by \eqref{relativearaki} when $\alset$ is not an algebra. We will call this quantity the ``Araki relative entropy'' and check whether it obeys the various properties that one expects of the relative entropy. We will follow the approach of \cite{narnhofer1998relative}.

\paragraph{\bf Proof of monotonicity of the Araki relative entropy \\}

We now show that the Araki relative entropy, \eqref{relativearaki}, is  monotonic under {\em projections} of the set of observables.  This means that if we take
\be
\blset \subset \alset,
\ee
where $\blset$ is also a linear space closed under the adjoint operation, then we find that
\be
\label{monotonicentr}
\relent{\blset}{\st}{\sst} \leq \relent{\alset}{\st}{\sst}.
\ee
To prove this relation we first note that the relative modular operator, when the set of observables is $\blset$, can be obtained from the relative modular operator defined on $\alset$ through
\be
\relmodsmall{\blset}{\st}{\sst} = P_{\blset} \relmod{\st}{\sst} P_{\blset},
\ee
where $P_{\blset}$ is a projector on $\hst$ that projects onto the little Hilbert space generated by $\blset$. Then we have
\be
\relent{\blset}{\st}{\sst} = -\langle \onest | \log\big[P_{\blset} \relmod{\st}{\sst} P_{\blset}\big] | \onest \rangle.
\ee
To compare this quantity with the original Araki relative entropy we adapt the argument of  \cite{narnhofer1998relative} and use the following identity.
Let $X$ be a bounded invertible matrix. Then, if $P$ is a projector, 
\be
\label{projinvident}
P X^{-1} P = P (P X P)^{-1} P + P X^{-1} (1 - P) \left[(1 - P) X^{-1} (1 - P) \right]^{-1} (1 - P) X^{-1} P.
\ee
What is important about this decomposition is that
when $X$ is a positive operator, we have
\be
\label{ineqprojinv}
P X^{-1} P \geq P (P X P)^{-1} P.
\ee
So, inverting and then projecting leads to a larger operator 
than projecting and inverting.

Since we can write (for any bounded and invertible operator)
\be
\log(X) = \lim_{\epsilon \rightarrow 0^+}\big(-\int_{0}^{1 \over \epsilon} {1 \over t + X} d t - \log(\epsilon)\big),
\ee
we see that the inequality \eqref{ineqprojinv} also implies that
\be
P \log( X ) P  \leq P \log (P X P) P.
\ee

In our case, we note that $P_{\blset} | \onest \rangle = |\onest \rangle$ and therefore we immediately see that
\be
 -\langle \onest | \log\big[ \relmod{\st}{\sst} \big] | \onest \rangle =  -\langle \onest | P_{\blset} \log\big[ \relmod{\st}{\sst} \big] P_{\blset} | \onest \rangle \geq -\langle \onest | \log\big[ P_{\blset} \relmod{\st}{\sst} P_{\blset} \big] | \onest \rangle,
\ee
which proves the relation \eqref{monotonicentr}.

We see that the nonnegativity of \eqref{relativearaki} also follows immediately. Since the trivial set of observables spanned by the identity operator, $\{1 \}$, is always a subset of $\alset$ and since \eqref{relativearaki} evaluates to $0$ for this set of observables we see that we always have $\relent{\alset}{\st}{\sst} \geq 0$.

\paragraph{\bf Failure of the Araki relative entropy to distinguish some states \\}
However, we now find that the Araki relative entropy cannot always distinguish states, when $\alset$ is {\em not} an algebra. The Araki relative entropy vanishes whenever {\em one-point} functions are equal
\be
\label{arakivanish}
\relent{\alset}{\st}{\sst} = 0 \iff \st(\al_i) = \sst(\al_i),
\ee
but this does {\em not} imply that $\st = \sst$ since we may still have $\st(\al_i \al_j) \neq \sst(\al_i \al_j)$ for some two-point functions.

To prove \eqref{arakivanish} we again follow \cite{narnhofer1998relative} and take $\blset$ to consist of {\em only} the identity operator (and its c-number multiples). Then $P_{\blset} |\al_i \rangle = \st(\al_i) |  \onest \rangle$. Moreover, with $X = t + \relmod{\st}{\sst}$ and applying \eqref{projinvident}, we see that 
\be
\relent{\alset}{\st}{\sst} = 0 \iff 
\int_0^{1 \over \epsilon} d t \langle \onest | P_{\blset} X^{-1} (1 - P_{\blset}) \left[(1 - P_{\blset}) X^{-1} (1 - P_{\blset}) \right]^{-1} (1 - P_{\blset}) X^{-1} P_{\blset} | \onest \rangle = 0.
\ee
Since $(1 - P_{\blset}) X^{-1} (1 - P_{\blset})$ is strictly positive, we see that 
\be
\relent{\alset}{\st}{\sst} = 0 \iff \langle \onest | P_{\blset} X^{-1}  (1 - P_{\blset}) X^{-1} P_{\blset} | \onest \rangle = 0, 
\ee
for all values of $t$. 
But then acting on $|\onest \rangle$ we must have
\be
(1 - P_{\blset}) X^{-1} | \onest \rangle = 0,
\ee
which can only happen if $X^{-1} | \onest \rangle \propto |\onest \rangle$ or multiplying both sides by $X$ if $X | \onest \rangle \propto | \onest \rangle$. The constant of proportionality can be set by sandwiching the state on the left with $\langle \onest |$ and we see that 
\be
\relent{\alset}{\st}{\sst} = 0 \iff \relmod{\st}{\sst} | \onest \rangle  = | \onest \rangle.
\ee
Sandwiching this expression with $\langle \al | $ for arbitrary $\al$ leads immediately to \eqref{arakivanish}.

To see a simple example of how non-identical states can have the same relative entropy consider a simple case where a basis for $\alset$ is formed by three elements $\{1, \al, \al^{\dagger} \}$ with $\al^{\dagger} = \al$.  Take the states
\be
\st(1) = 1; \quad \st(\al) = \st(\al^{\dagger}) = \st(\al^2) = \st((\al^{\dagger})^2) = 0; \quad \st(\al^{\dagger} \al) = \lambda_1; \quad \st(\al \al^{\dagger}) = \lambda_2; 
\ee
and a second state
\be
\sst(1) = 1; \quad \sst(\al) = \sst(\al^{\dagger}) = \sst(\al^2) = \sst((\al^{\dagger})^2) = 0; \quad \sst(\al^{\dagger} \al) = \mu_1; \quad \sst(\al \al^{\dagger}) = \mu_2.
\ee
Then it is clear that the states are not equal. But we see that, in the orthonormal basis $|\onest \rangle, {1 \over \sqrt{\lambda_1}} | \al \rangle, {1 \over \sqrt{\lambda_2}} |\al^{\dagger} \rangle$, the relative modular operator is given by
\be
\big[\relmod{\st}{\sst}\big]^O_{i j} = \begin{pmatrix}1 & 0 & 0 \\ 0 & {\mu_1 \over \lambda_1} & 0 \\ 0 & 0 & {\mu_2 \over \lambda_2} \end{pmatrix}.
\ee
It is clear that $\langle \onest | \log(\relmod{\st}{\sst}) | \onest \rangle = 0$, although the states are clearly unequal.

Note that if $\alset$ had been an algebra, this would not have happened. When $\alset$ is an algebra the state is completely characterized by its one-point functions. Indeed the difference between one- and higher-point functions is moot since all products of operators correspond to another operator in the algebra.  From \eqref{arakivanish} we see then that when $\alset$ is an algebra, the relative entropy vanishes only when the states are equal.

This is {\em not} in contradiction with the result on monotonicity which simply tells us that further projections will not reduce the entropy any further and keep it at 0.

We have not been able to find a distance measure that reduces to the relative entropy when $\alset$ is an algebra but also satisfies properties (1) -- (5) when $\alset$ is not an algebra. We should mention one other obvious generalizations of the relative entropy expressed in terms of modular operators.
\be
\label{relentnew}
\relentnew{\alset}{\st}{\sst} = {1 \over \tr\modop{\st}} \tr \left(\modop{\st} \left[ \log\big({\modop{\st} \over \tr (\modop{\st})}\big) - \log \big({\relmod{\st}{\sst} \over \tr(\relmod{\st}{\sst}) } \big) \right] \right),
\ee
where the traces are taken only over the {\em little Hilbert space} in an orthonormal basis. We can check that the expression \eqref{relentnew} reduces to the relative entropy when $\alset$ is an algebra by using the expression \eqref{relmodalg}. It also uniquely distinguishes states. This can be proved by recognizing that \eqref{relentnew} has the same {\em form} as the expression for the relative entropy in terms of two density matrices and therefore it can only vanish  when  $\modop{\st}/ \tr(\modop{\st})  =   \relmod{\st}{\sst}/\tr(\relmod{\st}{\sst})$. However, since $\langle \onest| \modop{\st} | \onest \rangle = \langle \onest| \relmod{\st}{\sst} | \onest \rangle$, we see that this in fact implies $\modop{\st} = \relmod{\st}{\sst}$.  Thus the distance measure in \eqref{relentnew} satisfies property \ref{proppos} . It is easy to see that also satisfies properties \ref{propbasisind} (basis-independence), \ref{propins} (insularity) and \ref{propaddit} (additivity).  However, this expression fails to satisfy property \ref{propmon} --- that of monotonicity.

We feel that such formulas deserve some further attention, and perhaps a refinement of \eqref{relentnew} or \eqref{relativearaki} can be engineered to satisfy all of properties \eqref{propbasisind} -- \eqref{propaddit} and also reduce to the relative entropy when $\alset$ is an algebra. 

We now turn to other measures of the distance that do satisfy the necessary properties but do not reduce to the relative entropy. 

\subsubsection{The normed entropy and other distance measures \label{secnormentropy}}
We now describe some measures of distance that satisfy all the properties \ref{propbasisind} to \ref{propins} including one measure --- the normed entropy --- that is also additive (property \ref{propaddit}).

The fundamental fact that we will exploit is \eqref{equalitycondition}: two states are equal only if the modular operator of one is equal to the relative modular operator to the other state. So, we will use the difference between these two operators as a measure of the distance between the two states. Accordingly, we define.
\be
\label{combopdef}
\combop = \modop{\st}^{-{1 \over 2}} \relmod{\st}{\sst} \modop{\st}^{-{1 \over 2}}.
\ee

If $\combop = 1$, then the modular and relative modular operators coincide, and so the states are the same. We want to define a measure of distance based on the spectrum of $\combop$.

One might imagine, at first glance, that this is simply a matter of comparing $\combop$ to the identity and using any of the standard matrix norms. However, we note that there is a tension between properties \eqref{propmon} (monotonicity) and \eqref{propins} (insularity). Insularity implies that the measure of distance should not change if we take $\combop \rightarrow \combop \otimes 1$ where $1$ denotes the identity matrix in an arbitrary number of dimensions. On the other hand, a simple way to satisfy  monotonicity is to use a norm that always decreases under $\combop \rightarrow P \combop P$, where $P$ is a projector.\footnote{Strictly speaking, this condition is sufficient, but not necessary, as we discuss in our proof of the monotonicity of the normed entropy.} To see the tension between these two requirements, consider the potential distance measure $\tr((\combop - 1)^2)$. This satisfies specificity and monotonicity. But clearly, it is not insular. We may attempt to correct for this, by dividing by the dimension:  ${1 \over \dima} \tr((\combop-1)^2)$. But then we lose monotonicity.

One way out is to use the {\em operator norm} $\|\combop\|$ and $\|\combop^{-1}\|$. Since $\combop$ is Hermitian and positive, these two norms simply correspond to the largest eigenvalue of $\combop$ and the inverse of its smallest eigenvalue. Note, by the result proved in \eqref{spectrumsame}, that with
\be
\sstcombop = \modop{\sst}^{-{1 \over 2}} \relmod{\sst}{\st} \modop{\sst}^{-{1 \over 2}},
\ee
we have
\be
\label{xinvy}
\|\combop^{-1}\| = \|\sstcombop\|.
\ee
Furthermore, it is easy to see that $\langle \onest | \combop | \onest \rangle = 1$ and also that ${}_{\sst} \langle \onest | \sstcombop | \onest \rangle_{\sst} = \onest$, and therefore the largest eigenvalue, $\|\combop\| \geq 1$, and the smallest eigenvalue, $\|\combop^{-1}\|^{-1} \leq 1$. The condition for states to be equal, which is that $\combop = 1$,  then becomes
\be
\st = \sst \iff \|\combop\| = {1 \over \|\combop^{-1}\|} = 1.
\ee
This just states that both the largest and smallest eigenvalues of $\combop$ become $1$.

The fact that any measure based on these operator norms will satisfy insularity (property \eqref{propins}) is obvious since
\be
\|\combop \otimes 1 \| = \| \combop \|; \quad \|\combop^{-1} \otimes 1 \| = \|\combop^{-1}\|.
\ee

Furthermore, these operator norms behave simply under contractions of $\alset$. 
Using the inequality \eqref{ineqprojinv} and using a similarity transformation to move the power of the modular operator entirely to the left we see that  for $\blset \subset \alset$ 
\be
 \| (P_{\blset} \modop{\st} P_{\blset}) ^{-1} P_{\blset} \relmod{\st}{\sst} P_{\blset} \| \leq   \| P_{\blset} ( \modop{\st} ) ^{-1} P_{\blset} \relmod{\st}{\sst} P_{\blset} \| \leq  \| \modop{\st}^{-1} \relmod{\st}{\sst} \|.
\ee
Therefore under a contraction of $\alset$, we see that $\|\combop\|$ {\em decreases}. By applying to same logic to the operator, $\sstcombop$, and using \eqref{xinvy} we see that $\|\combop^{-1}\|$ also {\em decreases} under a contraction of $\alset$. Therefore the smallest eigenvalue of $\combop$, $\|\combop^{-1}\|^{-1}$, increases.

So we conclude that any measure of distance specified by a nonnegative function of two variables, $\distfun: [1, \infty) \times [0,1] \rightarrow {\cal R}^+$
\be
\gendist{\alset}{\st}{\sst} = \distfun(\|\combop\|, {1 \over \|\combop^{-1}\|}),
\ee
will satisfy the conditions of specificity, insularity and monotonicity provided that the function satisfies 
\be 
\distfun(x,y) = 0 \iff x = y=1,
\ee
and 
\be
x_1 \leq x_2~\text{and}~ y_1 \geq y_2 \implies D(x_1, y_1) \leq  D(x_2, y_2).
\ee
In fact, the simplest choice for this function is just $\distfun(x,y) = x - y$ which translates into $\gendist{\alset}{\st}{\sst} = \|\combop\| - {1 \over \|\combop^{-1}\|}$.

If we additionally demand {\em additivity} (property \eqref{propaddit}) then this translates to the statement that $D(x_1 x_2, y_1 y_2) = D(x_1, y_1) + D(x_2, y_2)$ and a choice that satisfies this condition is  the ``normed entropy''. This is given by
\be
\label{opnormdef}
\snorm{\alset}{\st}{\sst}=\log \| \combop \|  +  \log \|  \combop^{-1} \|.
\ee
Using \eqref{xinvy} we can also write the normed entropy as $\snorm{\alset}{\st}{\sst} = \log \| \sstcombop \|  +  \log \|  \sstcombop^{-1} \|$.

The arguments above immediately tell us that the normed entropy is symmetric in its arguments, specific, monotonic, insular and additive. However, it is {\em not} finite. This can be easily seen by considering a pure state. For pure states, in the appropriate basis we see that several eigenvalues of the modular operator are zero. Thus, unless the second state also has the same set of zero-norm states in its little Hilbert space,  the normed entropy diverges when one uses it to measure the distance between one pure state and another state. 

However, if we do not insist on additivity, then it is not difficult to define a {\em finite} measure of distance by simply choosing another function $\distfun(x, y)$. For instance, we may take the measure of distance to be
\be
\label{chinormdef}
\chinorm{\alset}{\st}{\sst} = 1 - {2 \|\combop\| \over (1 + \|\combop\|)^2} -  {2 \|\combop^{-1}\| \over (1 + \|\combop^{-1}\|)^2}.
\ee
This measure of distance varies between $(0, 1)$ and vanishes only when $\|\combop\| = \|\combop^{-1}\| = 1$. Since $\|\combop\| \geq 1$ and $\|\combop^{-1}\| \geq 1$, this distance decreases with decreasing $\|\combop\|$ and decreasing  $\|\combop^{-1}\|$.  So, it is specific, insular and monotonic, and it is also clearly finite. However, it is not additive. The specific choice of the function in \eqref{chinormdef} is clearly not unique but is motivated by simplicity, and a desire to treat reciprocal eigenvalues in $\combop$ symmetrically.

\subsection{Summary of distance measures}
In this section, we first described how an appropriate measure of distance could be used to induce measures of entanglement 
on the set of states. 

We then discussed various notions of distance. For the convenience of the reader, the table below provides a summary of the properties of the various measures of distance that we have considered. As we see, only the normed entropy and the $\chi$-distance satisfy all of the necessary properties \ref{proppos}, \ref{propmon}, \ref{propins}. The normed entropy is additive  (property \eqref{propaddit}) but not finite whereas the $\chi$-distance is finite (property \eqref{propfinite}) but not additive. All the measures are independent of the basis chosen for $\alset$, and so they automatically satisfy property \eqref{propbasisind}, which we do not include in the table below.
\begin{center}
\resizebox{\textwidth}{!}{
\begin{tabular}{|>{\centering}m{1.5cm}|>{\centering}m{1.7cm}|>{\centering}m{2.1cm}|>{\centering}m{2.25cm}|>{\centering}m{2.1cm}|>{\centering}m{2.1cm}|>{\centering}m{2.1cm}|}
\hline
Symbol&Definition &Specificity  (Property \ref{proppos})&Monotonicity (Property \ref{propmon}) &Insularity  (Property \ref{propins}) &Additivity  (Property \ref{propaddit})&Finiteness  (Property \ref{propfinite})\tabularnewline\hline
$\relent{\alset}{\sst}{\st}$ & \eqref{relativearaki} & $\times$ & $\checkmark$ & $\checkmark$ &  $\checkmark$ &  $\times$\tabularnewline
$\relentnew{\alset}{\sst}{\st}$ & \eqref{relentnew} & $\checkmark$  & $\times$ & $\checkmark$ &  $\checkmark$ &  $\times$\tabularnewline
$\snorm{\alset}{\st}{\sst} $  & \eqref{opnormdef} &  $\checkmark $ & $\checkmark$ & $\checkmark$ &  $\checkmark$ &  $\times$\tabularnewline
$\chinorm{\alset}{\st}{\sst} $  & \eqref{chinormdef} &  $\checkmark $ & $\checkmark$ & $\checkmark$ &  $\times$ &  $\checkmark$ \tabularnewline
\hline
\end{tabular}
}
\end{center}

\section{Coarse and Fine-Grained Subregion Dualities in AdS/CFT \label{secadscft}}
We now describe some applications of our formalism to the formulation of subregion dualities in the  AdS/CFT correspondence.

We should clarify the relationship of our approach to the existing literature on holographic entanglement, following the  Ryu-Takayanagi conjecture \cite{Ryu:2006ef,Ryu:2006bv} and its generalization by Hubeny, Rangamani and Takayanagi \cite{Hubeny:2007xt}. The literature largely deals with the question of understanding {\em bulk geometry} from {\em boundary entanglement} \cite{VanRaamsdonk:2009ar,VanRaamsdonk:2010pw}. Here, our perspective is rather different. We are interested in studying quantum information measures directly from the point of view of the bulk gravitational theory. 

The question of bulk-entanglement has been studied only in the free-field limit \cite{Faulkner:2013ana,Jafferis:2015del}, where bulk quantities appear as one-loop corrections to the Ryu-Takayanagi formula.  (See, for example, \cite{Casini:2009sr,Casini:2014aia}, for nice reviews and calculations of quantum information measures in free quantum field theories.) Here, we take some initial steps towards clarifying the conceptual meaning of bulk quantum information measures and placing them in a framework where, in principle, there is no obstacle to turning on interactions in Newton's constant.

Accordingly, we consider a large-N conformal field theory that is dual to a bulk theory of quantum gravity in AdS$_{d+1}$. Since the boundary theory is at large $N$, we are in a regime where all curvature scales are large compared to the Planck scale and we additionally assume that the string coupling is small so that curvature scales are also widely separated from the string scale.  In this regime, the boundary theory has a natural set of ``simple operators'' called generalized free-field operators.  (In $N=4$ SYM, the generalized free-field operators are the single-trace operators at low dimension.) In the discussion below, we will denote such an operator by $O(t,x)$. To lighten the notation, we suppress tensor indices, and other indices required to distinguish between different generalized fields. 

We now consider the following questions:
\begin{enumerate}
\item
{\bf Coarse-Grained Subregion Duality Problem:} {\em Given a spacetime region $\reg$ on the boundary, is there a spacetime region $\bcoarse$ in the bulk so that all information about $\bcoarse$ can be obtained through measurements of low-order polynomials of generalized free-field operators in $\reg$?}. 
\item
{\bf Fine-Grained Subregion Duality Problem:} {\em Given a spacetime region $\reg$ on the boundary, is there a spacetime region $\bulkreg$ in the bulk so that all information about $\bulkreg$ can be obtained through arbitrary measurements in $\reg$?}. 
\end{enumerate}
In formulating the questions above, we  do {\em not} demand that $\reg$ is a causal diamond, but allow it to be an arbitrary region in spacetime. Moreover, we note that the fine-grained subregion duality problem is not reflexive. We do not demand that all information about $\reg$ can be obtained from $\bulkreg$.
 
We now examine the answer to these questions, and note the utility of our quantum information measures in testing these subregion dualities.

\subsection{\bf Coarse grained subregion dualities \label{seccoarsegrainedsubreg}}
To determine what information about the bulk may be obtained by measuring low-point correlation functions of generalized free-fields, we can organize these correlation functions as one and two-point functions of a set of operators that we can call $\alrcoarse$. 

More precisely, we need to do the following. We consider a lattice of points $(t_i, x_i) \in \reg$ and then consider the set of polynomials in these operators with an order limited by $n_{\text{coarse}}$.  We can choose $n_{\text{coarse}}$ to be any number parametrically separated from $N$.\footnote{In any actual calculation, it is also necessary to regulate the local generalized free-fields to turn them into bounded operators. This can be done, for example, by transforming to Fourier space to obtain the modes of these fields, and then cutting off the inverse Fourier transform by writing the ``local'' operators as a sum over a finite but large number of Fourier modes. We will discuss this further in forthcoming work.} A basis for this set is given by monomials in the generalized free fields.
\be
\label{alrcoarsedef}
\alrcoarse = \text{span~of}\{(\op(t_1, x_1) \ldots \op(t_n, x_n))\}, \quad n \leq n_{\text{coarse}}.
\ee

Note that $\alrcoarse$ is {\em not} an algebra. The  one- and two-point functions of operators in $\alrcoarse$ (which translate into correlators with up to 2 $n_{\text{coarse}}$ insertions of the generalized free-fields)  have information about a region in the bulk that we will call $\bcoarse$ --- this is a ``coarse-grained'' version of a subregion duality in AdS/CFT. We now describe the geometric structure of $\bcoarse$ more carefully.

First consider a single causal diamond inside $\reg$ that we call $\diam$. A causal diamond is specified by two points that are timelike to each other. If we call the later point, $P$, and the earlier point $P'$ then the causal diamond defined by these two points is given by
\be
\label{diampp}
\diam = \tilde{J}^{+}(P') \cap \tilde{J}^{-}(P),
\ee
where $\tilde{J}^{+}(P')$ and $\tilde{J}^{-}(P)$ denote the causal future of $P'$ and the causal past of $P$ respectively taken {\em only} on the boundary. 

Now, consider a scalar single-trace operator $O(t, x)$ with dimension $\Delta$ with  $(t, x) \in \reg$. This operator is dual to a bulk field $\phi(t, x, r)$ with mass given by $\Delta (\Delta - d) = m^2$. Here, we have introduced the coordinate  $r$ for the radial direction, with $r = \infty$ being the boundary and where the bulk metric diverges as $r^2$ as $r \rightarrow \infty$.  At large $N$, this bulk field obeys the bulk equation
\be
\label{phieom}
(\Box - m^2) \phi(t, x, r) = 0 + \Or[{1 \over N}],
\ee
where the corrections come from interaction terms. We can {\em solve} this bulk equation of motion
with boundary conditions specified on $\diam$ as
\be
\label{phibdry}
\lim_{z \rightarrow 0} r^{\Delta} \phi(t, x, r) = O(t, x), \quad \forall (t,x) \in \diam.
\ee
This solution simply leads to the standard HKLL kernel which expresses the bulk field as a function of its boundary values \cite{Hamilton:2005ju,Hamilton:2006az}.\footnote{As explained in \cite{Papadodimas:2012aq,Papadodimas:2013jku}, to avoid some of the difficulties outlined in  \cite{Bousso:2012mh,Rey:2014dpa} it is important to examine this kernel in momentum space rather than position space.}

When the bulk geometry is close enough to the AdS vacuum, the solution to \eqref{phieom}, subject to the boundary conditions \eqref{phibdry}, is valid within the {\em causal wedge} of $\diam$ in the bulk.   The causal wedge is defined as 
\be
\label{causdef}
\caus{\diam} = J^{+}(\diam) \cap J^{-}(\diam) ,
\ee
where now $J^{+}(\diam)$ and $J^{-}(\diam)$ denote the causal future and causal past  of a region but now {\em including} those points that lie in the bulk and are not on the boundary. For the causal diamond specified by \eqref{diampp}, note that we also have $\caus{\diam} = J^{+}(P') \cap J^{-}(P)$.  

The result that the bulk equations of motion can be solved within the causal wedge in a general background has not been established rigorously but we believe that it should be true because of the following simple physical intuition: if one draws the future and past null cone from any point in $\bcoarse$ then both these null cones intersect $\reg$. Therefore, operators in $\reg$ can ``sense'' the presence of any excitation within $\bcoarse$ purely through the classical propagation of gravity waves.

We can now use this to understand the coarse-grained dual of an arbitrary {\em spacetime} region on the boundary. First,  we consider all causal diamonds that fit inside $\reg$. Then it is clear that we have
\be
\reg = \bigcup_{\diam \subseteq \reg} \diam.
\ee
This is because any region $\reg$ can be completely tiled by boundary causal diamonds. Note that many of the diamonds within $\reg$ overlap.  
The causal wedge of any such diamond is defined as $\caus{\diam}$ as in \eqref{causdef}.

Since each causal diamond, $\diam$,  in $\reg$ is precisely dual to the set of operators in $\caus{D}$ in a coarse-grained sense, it is clear that the union of all causal diamonds is dual to the union of all causal wedges.  Therefore, the coarse-grained bulk dual of the region, $\reg$ that we denote by $\bcoarse$ is given by
\be
\label{coarsedual}
\begin{split}
&\reg \leftrightarrow \bcoarse,\quad  \text{(coarse~grained)} \\
&\bcoarse = \bigcup_{D \subseteq \reg} \caus{D}.
\end{split}
\ee
The only subtlety to keep in mind is that, in general,  
\be
\bcoarse \neq J^{+}(\reg) \cap J^{-}(\reg).
\ee
 This is because we do not necessarily have $J^{\pm}(\reg) \subseteq \reg$ for arbitrary spacetime regions that are not complete causal diamonds. 

We remind the reader that what the duality \eqref{coarsedual} means is that the simple operators that are part of $\alrcoarse$ give information about correlation functions of simple polynomials in bulk fields that live in $\bcoarse$. 
Note also that we derived \eqref{coarsedual} at large $N$. Even perturbatively, in $\Or[{1 \over N}]$ the notion of bulk locality is gauge-dependent in the bulk. However, we believe that it should be possible to choose a gauge so that \eqref{coarsedual} remains valid perturbatively in $\Or[{1 \over N}]$. This conjecture can be checked using our entanglement measures both at large $N$ and perturbatively in ${1 \over N}$ as described below.

\subsubsection{Information measures in coarse-grained subregion dualities}
The entanglement measures that we have described above are  tailor made to check the duality \eqref{coarsedual}. Consider two states $\st$ and $\sst$. We can probe these states both on the boundary, by using operators in $\alrcoarse$ and through simple bulk operators. Using a truncation procedure analogous to \eqref{alrcoarsedef}, we define the set of low-order polynomials in bulk fields to be $\albcoarse$.
Then we claim that 
\be
\label{coarseentmatch}
\snorm{\alrcoarse}{\st}{\sst} = \snorm{\albcoarse}{\st}{\sst}; \quad \chinorm{\alrcoarse}{\st}{\sst} = \chinorm{\albcoarse}{\st}{\sst}.
\ee

It would be tempting to conjecture that the full {\em spectrum} of the bulk and boundary relative modular operators matches. 
However, this is likely to depend  on the details of how precisely we regulate the bulk and the boundary theories. As a result \eqref{coarseentmatch} is likely to be more useful in practice.

Note that conventional entanglement measures {\em cannot} be applied to \eqref{coarsedual} (except at infinite $N$) since neither the bulk, nor the boundary set of operators form an algebra. 

The simplest example of \eqref{coarsedual} is where we restrict $\alrcoarse$ to consist of just {\em single insertions} of single-trace operators. Then our entanglement measures are sensitive to two-point functions, and this is already sufficient to check \eqref{coarsedual} at large $N$.  At the next level, we can restrict $\alrcoarse$ to consist of single-insertions of either single-trace or double-trace operators and consider the first non-trivial power of $\Or[{1 \over N}]$ in all correlation functions. Then \eqref{coarseentmatch} already becomes sensitive to perturbative corrections but can still be calculated at least in some simple examples. This can be used to check \eqref{coarsedual}. We will comment further on this in forthcoming work.

\begin{figure}
    \centering
    \begin{subfigure}[b]{0.29\textwidth}
        \includegraphics[width=\textwidth]{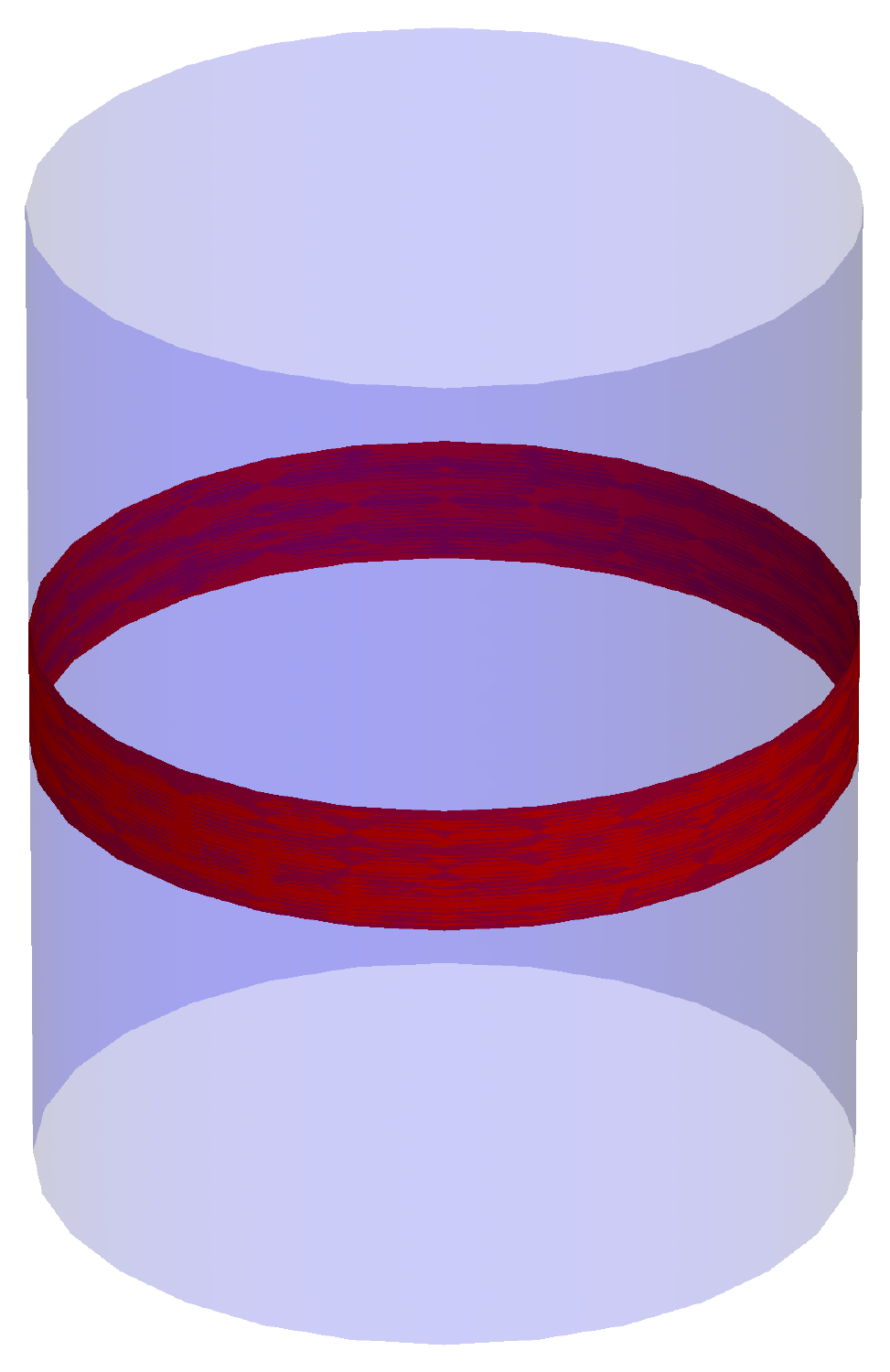}
        \caption{\em A time-band on the AdS boundary}
        \label{fig:band}
    \end{subfigure}
    \quad           \begin{subfigure}[b]{0.29\textwidth}
        \includegraphics[width=\textwidth]{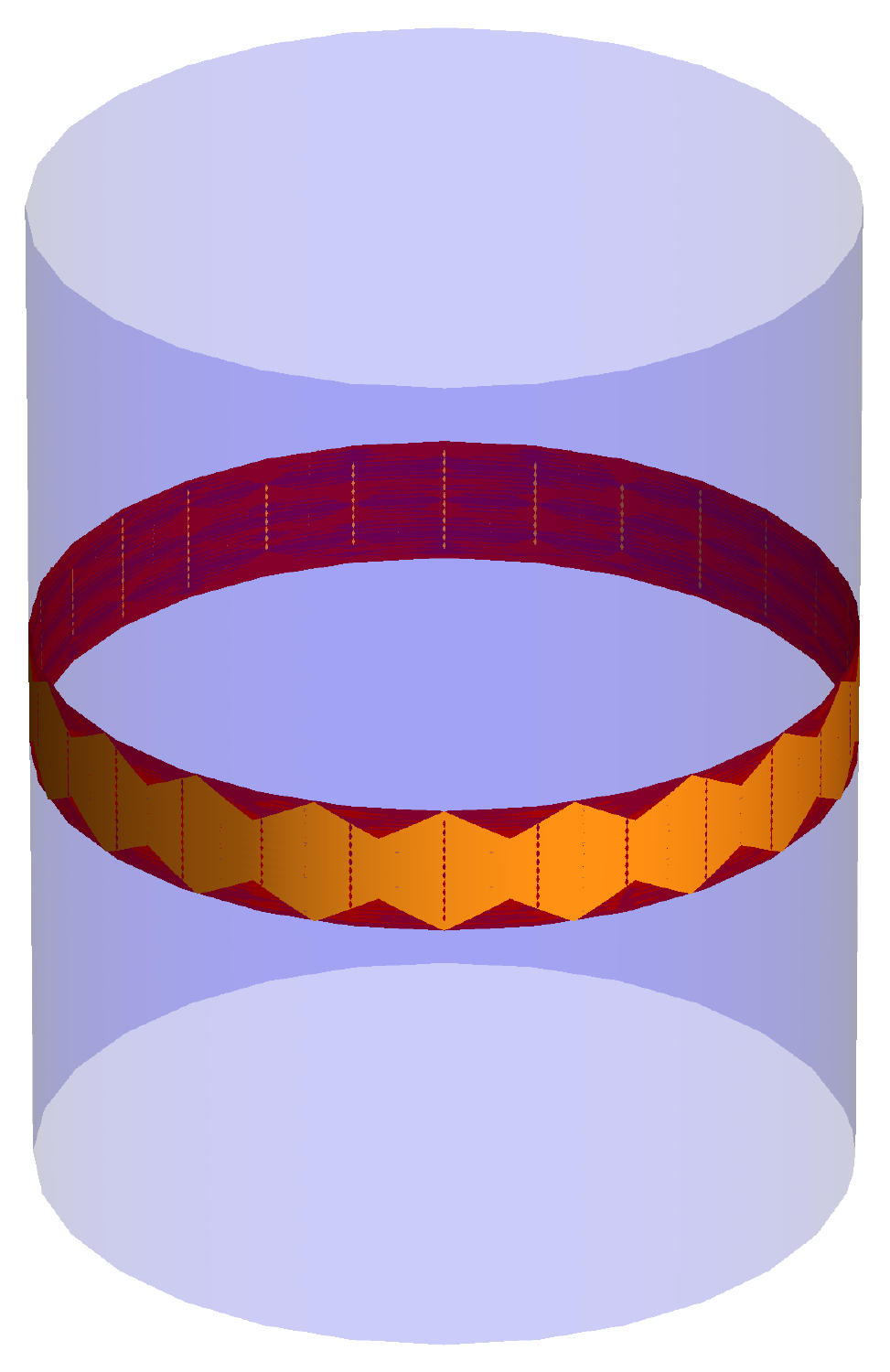}
        \caption{\em Foliating the band with overlapping diamonds}
        \label{fig:diamonds}
    \end{subfigure}
    \quad         \begin{subfigure}[b]{0.29\textwidth}
        \includegraphics[width=\textwidth]{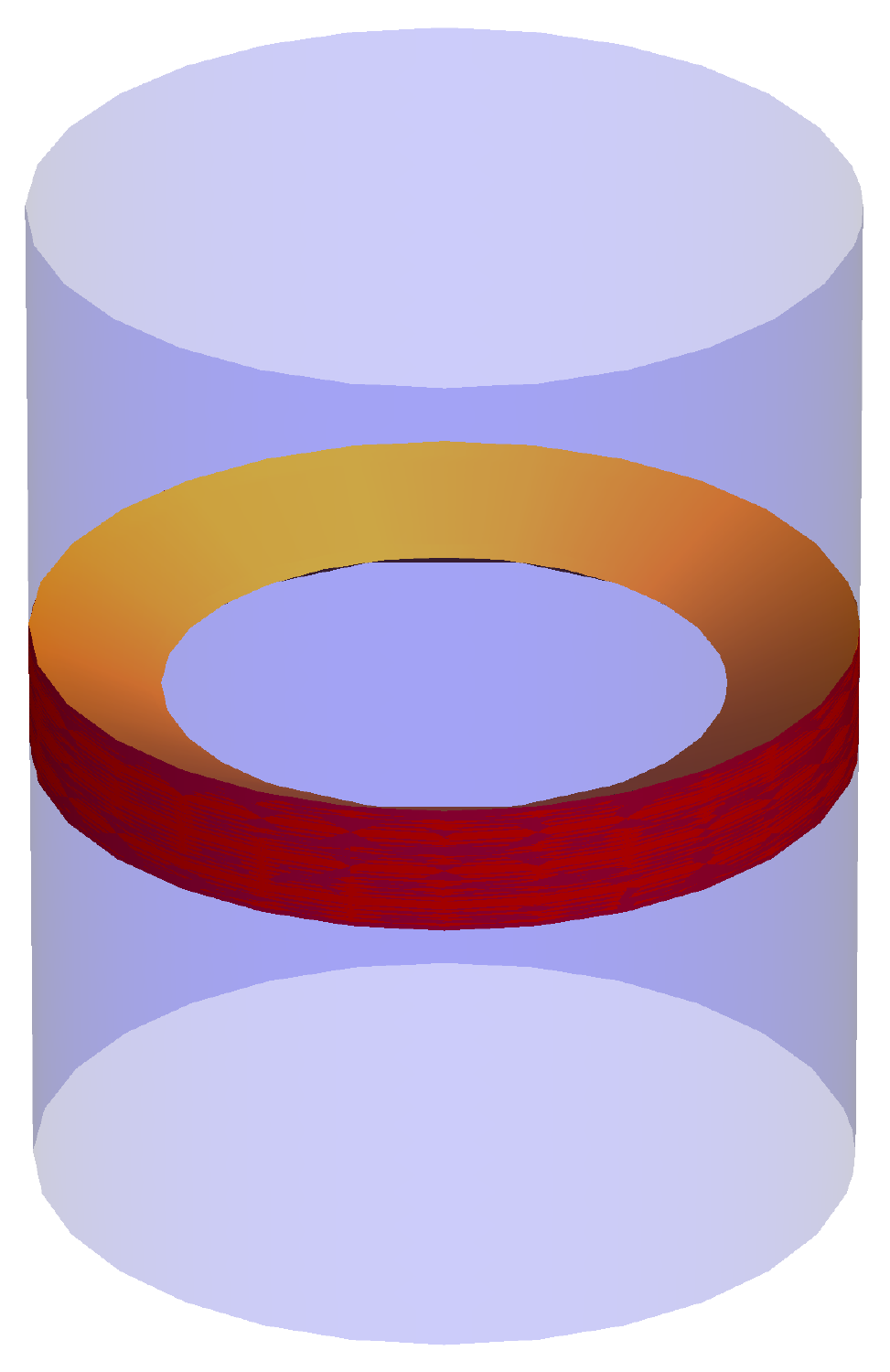}
        \caption{\em The coarse-grained dual bulk region}
        \label{fig:bulkdual}
    \end{subfigure}
    \caption{\em The coarse-grained dual of a time-band obtained by following the procedure summarized in \eqref{coarsedual}.}\label{fig:timeband}
\end{figure}

\paragraph{\bf The boundary time band: an example of coarse-grained subregion dualities \\}
As an example of a subregion duality of the kind that we are interested in, we revisit an example first considered in \cite{Banerjee:2016mhh}. Consider global AdS$_{d+1}$ with metric $ds^2 = -(1 + r^2) dt^2 + {d r^2 \over 1 + r^2} + r^2 d \Omega_{d-1}^2$ and a time band on the boundary that covers the entire sphere but extends in time from ${-T \over 2}$ to ${T \over 2}$,  where $T < \pi$. (For more details we refer the reader to \cite{Banerjee:2016mhh}.) Then this time-band can be tiled with multiple overlapping diamonds as shown in Figure \ref{fig:diamonds}. The coarse-grained dual of this time band is the complement of a bulk causal diamond as shown in figure \ref{fig:bulkdual} that is bounded by the light-sheets $r =  \cot\big({T \over 2} \pm t\big)$.  As we will see in the next subsection, the fine-grained dual of this same region is all of AdS!

\subsection{\bf Fine grained subregion dualities: \label{secfinegrainedsubreg}}
We will now enlarge the set of operators from $\alrcoarse$ to a larger set that we call $\alrfine$. If we denote the set of {\em all} boundary operators within $\reg$, by $\alr$, then the set $\alrfine \subset \alr$. Nevertheless we show that by considering the set of operators $\alrfine$,  we can obtain information about a bulk region $\bfine$ that is, in general, larger than $\bcoarse$. The interesting part of this fine-grained duality $\reg \leftrightarrow \bfine$ is that it may violate bulk causality in the sense that there may be points in $\bfine$ that are {\em not} causally connected to $\reg$.

To define, $\alrfine$, we again consider the set of generalized free-fields. One light generalized free-field that must exist in any conformal field theory is the stress tensor $T^{\mu \nu}$.  For notational simplicity we assume that the boundary is either $S^{d-1} \times \reg$ or $R^{d-1} \times R$ so that ${\partial \over \partial t}$ is a Killing vector on the geometry. We then consider the set of all spacelike slices, $\cauch$, within the region $\reg$ and on each such slice we define 
\be
\hamilt{\cauch} = \int_{\cauch} T^{t \mu} d \Sigma_{\mu},
\ee
where $d \Sigma_{\mu} = \sqrt{h} n_{\mu} d^{d-1} x$ and $n_{\mu}$ is the future-directed unit normal to $\cauch$ and $h$ is the induced metric on $\cauch$. Note that if $\cauch$ had been a complete Cauchy slice then $\hamilt{\cauch}$ would have reduced to the Hamiltonian, $H$,  of the theory. 

Now, denote the {\em causal completion} of $\cauch$ on the {\em boundary} by $\causalcomp{\cauch}$. The causal completion is defined as follows. We consider all points that are spacelike to $\cauch$ and denote them by $\cauch'$. Then $\causalcomp{\cauch}$ is the set of all points that are spacelike to $\cauch'$. 

Now, since the boundary theory is exactly causal, for {\em any} Heisenberg operator on the boundary theory that is localized within $\causalcomp{\cauch}$ we have
\be
{d \op(t, x) \over d t} = i [H, \op(t,x)] = i [\hamilt{\cauch}, \op(t, x)].
\ee
This is because $\op(t,x)$ commutes with the Hamiltonian density outside $\cauch$.

Therefore all operators within $\causalcomp{\cauch}$ can be obtained through time-evolution with $\hamilt{\cauch}$. In particular an operator at a point $(t + \tau, x)$ can be written as
\be
\label{hamiltwithincausalcomp}
\op(t + \tau, x) = e^{i \hamilt{\cauch} \tau} \op(t, x) e^{-i \hamilt{\cauch} \tau}, \quad (t + \tau, x) \in \causalcomp{\cauch}; \quad (t, x) \in \cauch.
\ee

In fact \eqref{hamiltwithincausalcomp} can be well approximated by a high order polynomial in simple operators by simply writing
\be
\label{hamiltwithincausalcompapprox}
\op(t + \tau, x) \approx \sum_{n=0}^{\ncut} {i^n \tau^n \over n!} \underbrace{[\hamilt{\cauch}, \ldots [\hamilt{\cauch}, \op(t, x)]]}_{n~\text{times}}.
\ee
If we cut this series off using $\ncut = \Or[N]$ terms, and if we also ensure that the operators \eqref{hamiltwithincausalcompapprox} are only used within correlators where the minimum separation between points is parametrically larger than $\Or[{1 \over N}]$ then we see that \eqref{hamiltwithincausalcompapprox} gives an excellent approximation to \eqref{hamiltwithincausalcomp}.

We now define $\alrfine$ to be the set of simple polynomials in the operators \eqref{hamiltwithincausalcomp}. More precisely, we take
\be
\alrfine = \text{span~of}\{(\op(t_1 + \tau_1, x_1) \ldots \op(t_n + \tau_n, x_n))\}, \quad  (t_i + \tau_i, x_i) \in \causalcomp{\cauch}, \quad \cauch \subset \reg; \quad n \leq n_{\text{coarse}}.
\ee
Note that, using \eqref{hamiltwithincausalcompapprox},  $\alrfine$ can be thought of as a combination of the simple polynomials in $\alrcoarse$ and a set of simple polynomials of a {\em very specific set} of complicated polynomials in the elements of $\alrcoarse$. 

The bulk dual to the set of operators, $\alrfine$ follows immediately from the previous duality. The set of simple operators in each causal completion $\causalcomp{\cauch}$ is dual to a set of simple operators in the bulk region given by $\caus{\causalcomp{\cauch}}$, which is defined just as in the previous subsection. Therefore if we define 
\be
\label{finegrained}
\bfine = \bigcup_{\cauch \subset \reg} \caus{\causalcomp{\cauch}},
\ee
then the set of operators in $\alrfine$ is dual to a set of simple bulk operators that live on $\bfine$. We denote this set by $\albfine$.

It is clear that  the boundary region $\reg$ must probe $\bfine$ in any theory of quantum gravity in anti-de Sitter space. This is because all that we have used to establish the form of $\bfine$ in \eqref{finegrained} is the fact that the canonical Hamiltonian is a boundary term \cite{DeWitt:1967yk,Regge:1974zd,Arnowitt:1962hi} and that asymptotic operators commute exactly at spacelike separation.

As we mentioned above, the curious part of  the duality $\alrfine \leftrightarrow \albfine$ is that it may violate bulk causality in the sense that there may be points in $\bfine$ that are not causally connected to $\reg$. For instance consider the time-band shown in Figure \ref{fig:timeband}. Then one spacelike slice that lives within this time band is simply the slice with $t = 0$ on the boundary. The causal completion of this slice is the entire boundary. Therefore the region $\bfine$ corresponding to  the time-band is all of AdS! This is in sharp contrast to the coarse-grained bulk dual region, $\bcoarse$, which is just the region shown in Figure \ref{fig:bulkdual}.

This duality also gives us an example where the set of approximately-local operators in a region may not form an algebra. Note that the elements of $\albfine$ can also be generated by taking complicated polynomials of $\albcoarse$. But now we see that these complicated polynomials should be understood as simple field operators in the region $\bfine$ that, in general, is larger than $\bcoarse$. Therefore complicated polynomials of approximately-local operators in $\bcoarse$ do not remain meaningfully confined to $\bcoarse$.

\subsubsection{Information measures in fine-grained subregion dualities}
Note that, as we have defined it above, $\alrfine$ is also {\em not} an algebra. However, our quantum information measures can still be applied to the duality above.  This duality implies the equality of our distance measures between two states, when they are probed by $\alrfine$ and $\albfine$. 
\be
\snorm{\alrfine}{\st}{\sst} = \snorm{\albfine}{\st}{\sst}; \quad \chinorm{\alrfine}{\st}{\sst} = \chinorm{\albfine}{\st}{\sst}.
\ee

Note, however, that we still have $\alrfine \subset \alr$. Therefore, by the monotonicity of the distance measures we also have
\be
\label{bdrgtbulk}
\snorm{\alr}{\st}{\sst} \geq \snorm{\albfine}{\st}{\sst}; \quad \chinorm{\alr}{\st}{\sst} \geq \chinorm{\albfine}{\st}{\sst}.
\ee

As we will see below, this may correspond to the fact that the region $\bfine$ is still smaller than the full region dual to $\alr$.

\paragraph{\bf Entanglement wedges \\}
Before, we close this section, we should note that it is, of course, possible to extend the sets $\alrcoarse$ and $\alrfine$ into algebras, simply by taking the set of all polynomials in the generalized free-fields. This leads us to the set of {\em all} operators in the region $\reg$, which we have called $\alr$.

The bulk dual to $\alr$ has been studied extensively in the literature, and it is generally believed that, for each spacelike slice within $\reg$, the bulk dual corresponds to the {\em entanglement wedge} of the slice  \cite{Headrick:2014cta}. The entanglement wedge of a boundary spacelike slice is the bulk region that is causally determined by data on the Ryu-Takayanagi surface that ends on the boundary slice.  The union of such entanglement wedges should then give us the complete bulk dual to $\reg$, which we denote by $\bfinefine$.  The strongest evidence for this claim is that the relative entropy between two states evaluated on $\bfinefine$ is equal to the relative entropy evaluated on $\reg$ \cite{Jafferis:2015del,Dong:2016eik}. We do not know of any direct proof of this conjecture although if one {\em assumes} that operators in $\bfinefine$ are dual to operators in $\reg$ then it is possible to write down a formula relating the two sets of operators \cite{Faulkner:2017vdd}. 

Note that, in general, the region specified in \eqref{finegrained} is a {\em subset} of the entanglement wedge. Thus, the entanglement-wedge proposal suggests that locality is violated even more strongly than is suggested by \eqref{finegrained}.   The fact that $\bfine \subseteq \bfinefine$  is just the {\em geometric analogue}  of \eqref{bdrgtbulk}.

We note that if one is just interested in studying the duality $\reg \leftrightarrow \bfinefine$, which holds when we consider the set of all operators in $\reg$ then, since $\alr$ is an algebra, conventional quantum information measures work well, and our quantum information measures do not have any particular role to play.

We also note that the bulk region $\bfinefine$ is an example of a region, where the set of local operators can be completed to form an algebra. This algebra is just $\alr$.  If we take arbitrary products of local operators in the region, $\bfinefine$, this gives us other elements of $\alr$ but does not allow us to expand our knowledge about the bulk to any region larger than $\bfinefine$.

\section{Conclusion}
In this paper, we have described measures of quantum information that are applicable when we can only probe a system with a limited number of observables, and when the space spanned by these observables, $\alset$,  does not close to form a  von Neumann algebra.   We believe that these measures are particularly relevant in quantum gravity where, for physical reasons, the set of approximately localized operators in a region may not close to form an algebra. A corollary to this fact is that the Hilbert space of gravity does not factorize into a Hilbert space associated with a region, and its complement.   

However, even for simpler systems that do not contain gravity,  the set of accessible observables may not be closed under multiplication due to physical or experimental limitations. We believe that the information measures we have defined here serve as a natural generalization of conventional information measures and may be  useful in a study of such systems.

One of the central conclusions of this paper is that the objects that deserve attention in this situation are the {\em modular} and the {\em relative modular} operators.  When $\alset$ is an algebra,  these operators can be written in terms of the density matrix associated with the state. However, these operators can be defined more generally, and their spectrum gives us a characterization of the state that is invariant under general linear transformations of the basis used for $\alset$.
Moreover in sections \ref{secmodpairing}, \ref{secrelmodinv}, \ref{secspectrelationship}, we showed that  several properties of these spectra, which are obvious when $\alset$ is an algebra, continue to hold even when $\alset$ is not an algebra.

A key quantum information measure is the ``distance'' between states. An appropriate notion of distance, which obeys various properties that we reviewed in section \ref{secdistmeasures}, can then be used to define notions of bipartite or multipartite entanglement. When $\alset$ is an algebra,  the relative entropy is commonly used to describe the distance between states. However, when $\alset$ is not an algebra, the simplest generalization of the relative entropy fails the test of {\em specificity}:  the relative entropy may vanish even when two states are not equal. 

We proceeded to describe an entire class of distance measures in section \ref{secnormentropy} which did meet all of our desired properties. We focused on two of these measures --- the normed entropy, which was additive,  and the $\chi$-distance, which was finite.  These distance measures rely on the operator norm of a combination of the relative modular operator and the modular operator. We used these distance measures to describe notions of entanglement that measured only quantum correlations between systems, were invariant under local unitary transformations and decreased or remained constant under LOCC operations.

It is clear that there are several directions for further work. The first set of questions is purely information-theoretic. First, the distance measures we have defined are only a subset of a large class of distance measure that can be defined through the spectra of the modular and relative modular operators.  We believe that other possible distance measures should also be investigated and classified. Second, our measure of entanglement is defined by calculating the minimum distance between a state and the set of separable states. As we mentioned above determining the closest separable state is, even numerically, a difficult task.  So, it would be nice to understand if simpler measures of entanglement can be defined.

A third interesting question is whether we can define notions of distance that rely on a trace, rather than the operator-norm. As we explained above, the reason for using the operator-norm was to resolve the tension between the property of ``insularity'' and the property of ``monotonicity.'' 
In the case where $\alset$ is an algebra, trace-based measures like the relative entropy satisfy both these required properties since they are defined using density matrices, which are bounded and have unit trace. However, the modular operator does not share this property.  Nevertheless, it is not clear that this difficulty is insurmountable and this question deserves further attention particularly since trace-based measures may be ``smoother'' than measures based on the operator norm. 

Finally, an important property of entanglement is that it is monogamous. How, precisely, is the monogamy of entanglement reflected in these entanglement measures? These questions are particularly relevant, when these measures of entanglement are applied to theories of gravity since the monogamy of entanglement plays a significant role in precise formulations of the information paradox.

Another question,  which  is relevant while applying these measures to quantum field theory or gravity has to do with the UV-sensitivity of our measures. In our discussions above, we assumed  that all the relevant operators  had been regulated, so that they could be treated as bounded operators in a finite Hilbert space. However, in quantum field theories, it is clear that various quantities, such as the spectrum of the modular and the relative modular operator, may be sensitive to the UV-cutoff. We believe that the distance measures defined in section \ref{secdistmeasures} should be UV-safe, but it is clearly important to investigate this further.

\begin{wrapfigure}[15]{r}{0.32\textwidth}
\includegraphics[width=0.3\textwidth]{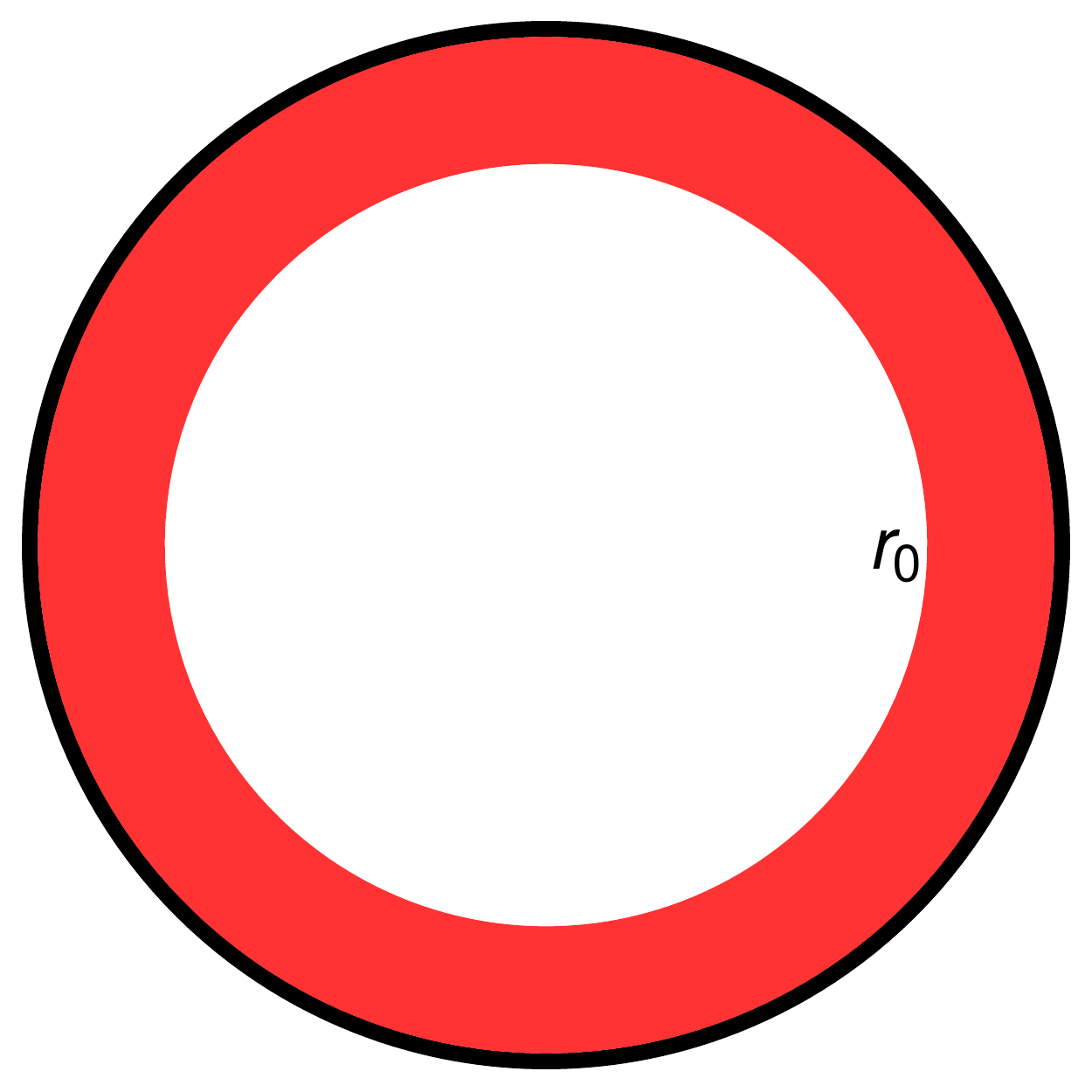}
\caption{\em A cross section of global AdS at constant time with the annular region $r > r_0$ marked in red. \label{annulusads}}
\end{wrapfigure}
There are several natural questions 
about entanglement in gravity, for which our formalism seems relevant.  For example, in the AdS/CFT correspondence,  a natural question is as follows. Consider the ``annular'' subregion of global AdS, at constant time,  with $r > r_0,$ where $r$ is the radial coordinate and $r_0$ is some cutoff.  The bulk dual of the time-band that is shown in Figure \ref{fig:bulkdual} is just the bulk causal completion of this annulus.  The complement of this region is the ``disk'' with $r < r_0$.   We would then like to analyze the entanglement  between the region $r > r_0$ and the region $r < r_0$. (See figure \ref{annulusads}.)

Within the conventional setting, this question can only be addressed at infinite $N$, where the effects of gravity are irrelevant. This is because, as we discussed in section \ref{secfinegrainedsubreg}, if we consider {\em all} operators in the annular region, and since the annular region includes a slice of the boundary at constant time, then this set is the complete set of operators in the theory. So all operators in the region $r < r_0$ are already included in the region $r > r_0$!

However, our measures of entanglement 
can be applied to this question, if we simply use the coarse-grained set of bulk and boundary operators described in section \ref{seccoarsegrainedsubreg}. Moreover, if we restrict the set of coarse-grained observables to polynomials that contain only a small number of insertions of generalized free-fields, then we believe that our measures should be computable numerically, at least to the first non-trivial order in ${1 \over N}$. We believe that this is an interesting problem, and we hope to comment further on this in forthcoming work.

\section*{Acknowledgments}
We would like to thank Ronak Soni, Sandip Trivedi,  and M.V. Vishal for collaboration in the early stages of this work and for several helpful discussions. We are grateful to Abhishek Agarwal, Jared Kaplan, R. Loganayagam, Kyriakos Papadodimas, Shiroman Prakash, Mukund Rangamani,  the participants of the Bangalore Area String Meeting (2017) and the participants of the National Strings Meeting (2017) for useful discussions. We are grateful to Arun Pati for comments on a draft of this paper. S.R. is partially supported by the Swarnajayanti Fellowship of the Department of Science and Technology (India).

\appendix
\section*{Appendix}
\section{Some Sample Calculations \label{appsample}}
In this appendix, we provide a few sample calculations using our formalism. Our objective is to help the reader understand the formalism, but also to show how this formalism allows for easy numerical computations of quantum information measures. We start with a physical example of a spin-chain system, and then show how an arbitrary finite-dimensional system can be analyzed.

\subsection{Distance measures in the spin chain}
We consider a spin-chain with $N$ spin-1/2 particles. These spins are acted on by a set of  Pauli spin operators denoted by $\sigma^{i}_{\alpha}$. Here $i \in [1,\cdots, N]$ labels the operators are different sites and  $\alpha \in [1,2,3]$ labels the different Pauli operators at a given site. These operators obey the usual commutation relations $[\sigma^i_{\alpha}, \sigma^j_{\beta}] = 2 i \delta^{i j} \epsilon_{\alpha \beta \gamma} \sigma^{i}_{\gamma}$. The dimension of the full Hilbert space is  $2^N$. 
\begin{figure}[!h]
\begin{center}
\begin{subfigure}[t]{0.45\textwidth}
\includegraphics[width=\textwidth]{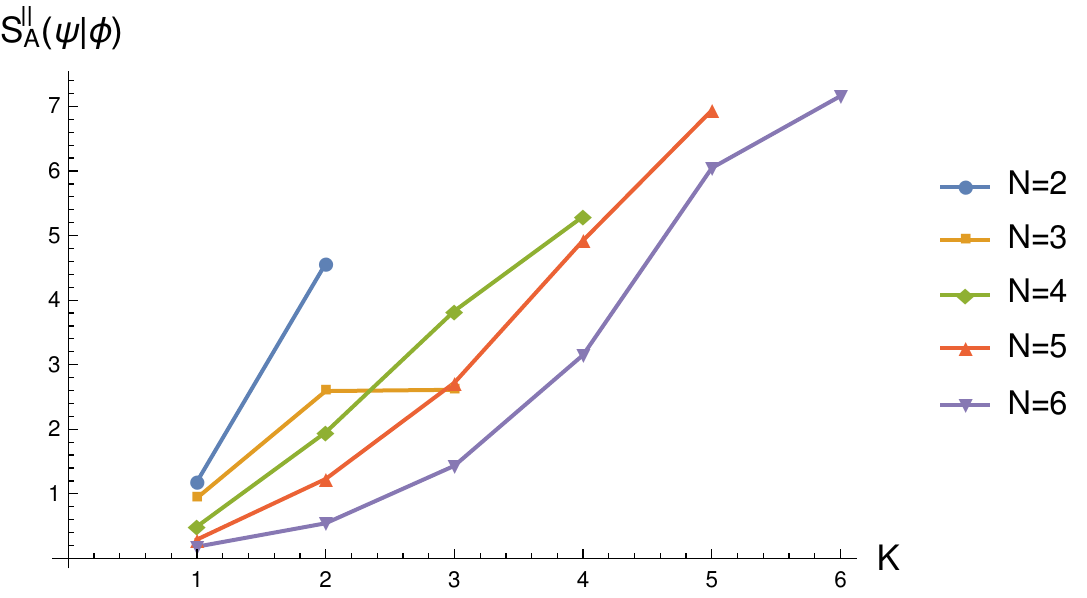}
\end{subfigure}
\hspace{0.05\textwidth}
\begin{subfigure}[t]{0.45\textwidth}
\includegraphics[width=\textwidth]{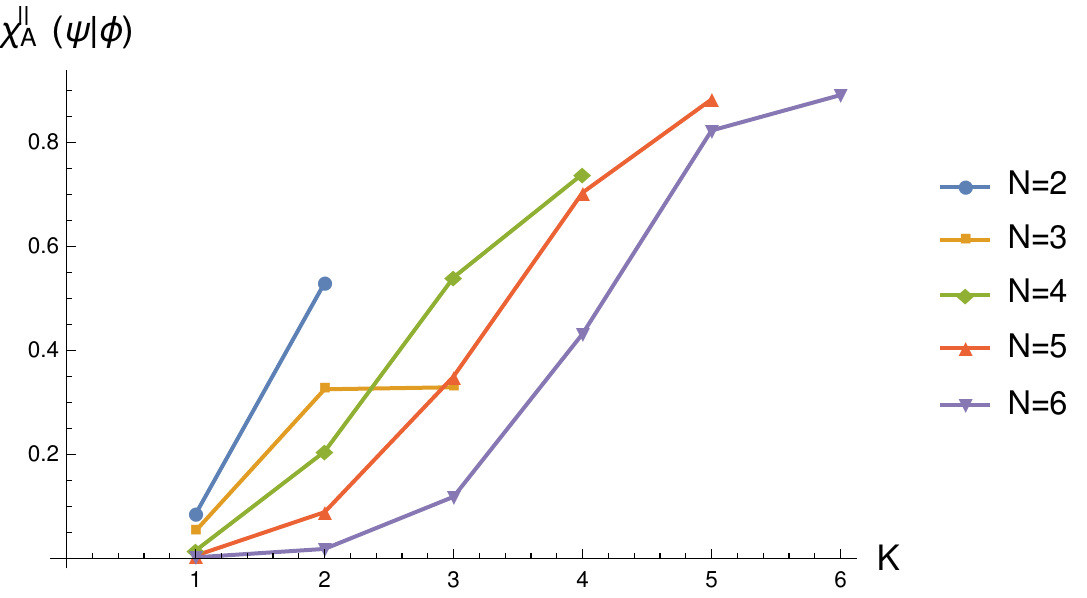}
\label{chinormspin}
\end{subfigure}
\caption{\em A plot of $\snorm{\alset}{\st}{\sst}$ (left) and $\chinorm{\alset}{\st}{\sst}$ for a system with $N$ spins probed with polynomials of order at most $K$ in the individual spin operators. \label{figqispin}}
\end{center}
\end{figure}

We take the set of accessible operators to be polynomials in these elementary spin operators with up to $K$-insertions.  Thus $\alset$ is given by
\begin{equation}
\label{spinalsetbasis}
\alset = \text{span~of~}\{\sigma_{\alpha_{1}}^{i_{1}}\sigma_{\alpha_{2}}^{i_{2}}\cdots\sigma_{\alpha_{n}}^{i_{n}}\},\quad ~ n \leq K. 
\end{equation}
The dimension of this set is given by
\begin{equation}
\dima = \sum_{j=0}^{K}\binom{N} {j}3^{j}.
\end{equation}
A basis of $\alset$ is given by the monomials displayed in \eqref{spinalsetbasis} and we denote them by $\al_1 \ldots \al_{\dima}$. Note that all these operators are Hermitian.

We take our states, $\st, \sst$ to be general mixed states in this system. They can be specified by $2^N \times 2^N$ density matrices and we denote these density matrices by $\rho$ and $\sigma$ respectively. We generate these density matrices as random unit-trace operators that are positive and self-adjoint. We then calculate the following matrices that are defined in the text. (Here, we have implemented the simplification that $\al_i^{\dagger} = \al_i$.)
\be
g_{i j} = \tr(\rho \al_i \al_j); \quad (\modop{\st})_{i j} = \tr(\rho \al_j \al_i); \quad (\relmod{\st}{\sst})_{i j} = \tr(\sigma \al_j \al_i).
\ee

Then, it is not difficult to see that the spectrum of the matrix $\combop$ defined in \eqref{combopdef} can be calculated simply by calculating the spectrum of the matrix
\be
\spect(\combop) = \spect\left(\modop{\st}^{-1} \cdot \relmod{\st}{\sst} \right),
\ee
where  $\modop{\st}^{-1}$ just denote the inverse of the modular operator and we have suppressed the matrix indices to lighten the notation. This is because the elements of $\combop$ in an {\em orthonormal} basis are related to the matrix on the right hand side of the equation above by a similarity transformation. The largest and smallest eigenvalues in this spectrum  give us $||\combop||$ and $1/||\combop^{-1}||$. 

To compute this spectrum is straightforward in principle, but quickly becomes computationally expensive for large $N$. All the matrices above are $\dima \times \dima$ sized matrices. In the table below, we give the value of $\dima$ for various values of $K$ and $N$. Note that we must have $K \leq N$ and so entries with $K > N$ are omitted.
\begin{center}
\begin{tabular}{|l|l|l|l|l|l|l|}
\hline
&K=1&K=2&K=3&K=4&K=5&K=6\\ \hline
N=2&7&16&&&&\\
N=3&10&37&64&&&\\
N=4&13&67&175&256&&\\
N=5&16&106&376&781&1024&\\
N=6&19&154&694&1909&3367&4096\\ \hline
\end{tabular}
\end{center}

Figure \ref{figqispin} shows the normed entropy and the $\chi$-distance for various values of $N$ and $K$ computed numerically for randomly generated states.

\subsection{Distance measures for random  matrices}
While our example in the previous section was physically motivated it is also possible to compute our quantum information measures when the matrix of correlation functions is given by a random matrix.
\begin{figure}[!h]
\begin{center}
\begin{subfigure}[t]{0.45\textwidth}
\includegraphics[width=\textwidth]{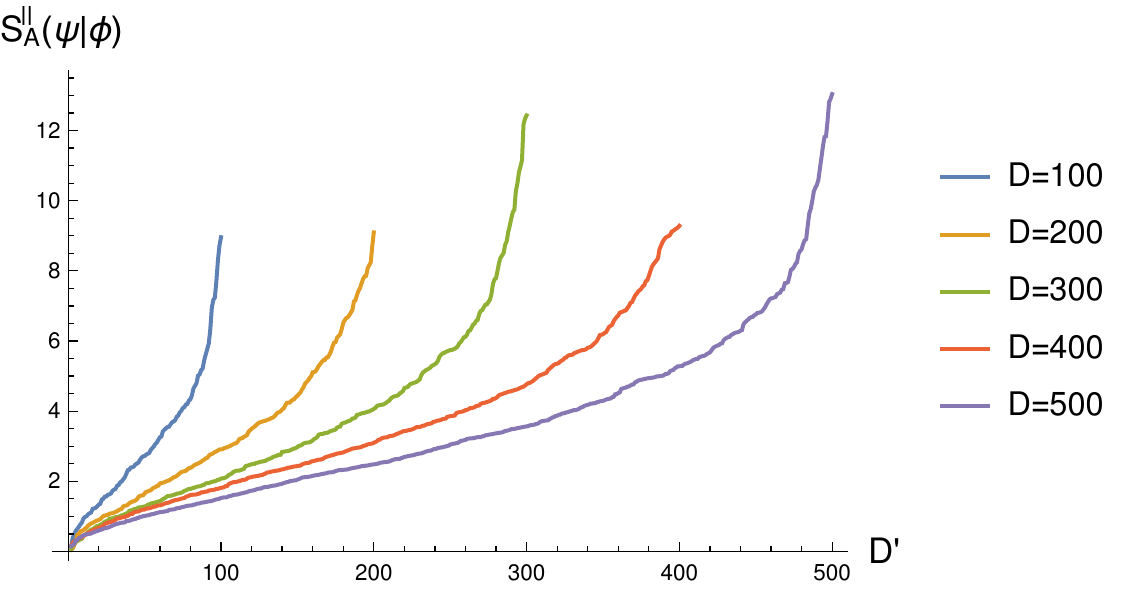}
\end{subfigure}
\hspace{0.05\textwidth}
\begin{subfigure}[t]{0.45\textwidth}
\includegraphics[width=\textwidth]{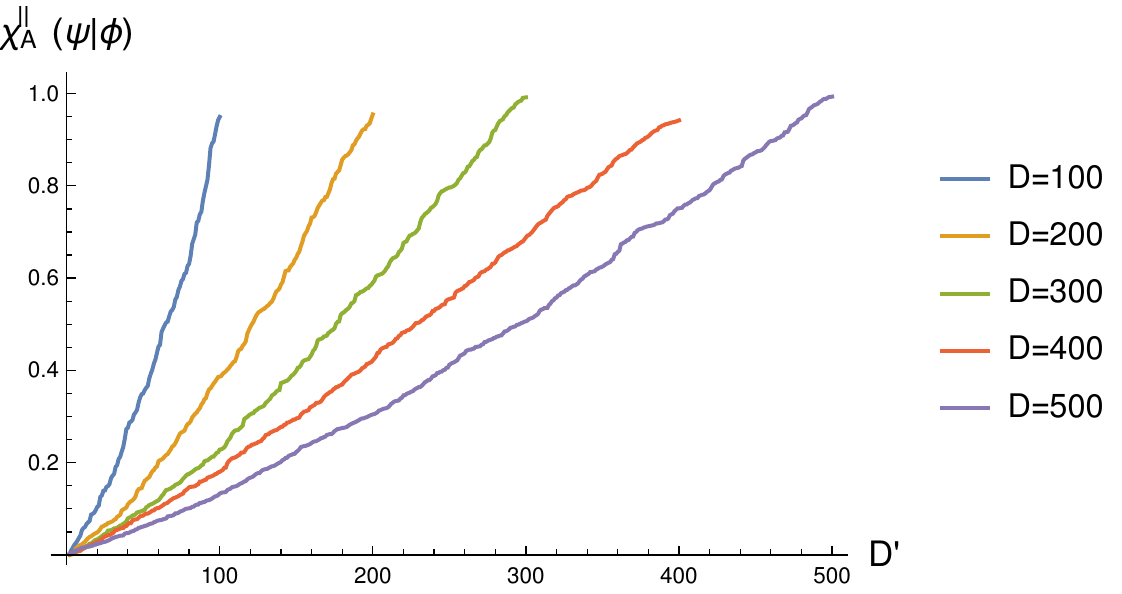}
\label{chinormspin}
\end{subfigure}
\caption{\em A plot of $\snorm{\blset}{\st}{\sst}$ (left) and $\chinorm{\blset}{\st}{\sst}$ for a system that originally has $\dima$ operators probed with a $\dima'$-dimensional  subset of these operators. (Note that successive values of $\dima'$ are joined leading to the impression of a continuous curve.) \label{figqimat}}
\end{center}
\end{figure}

More precisely, we take $\alset$ to consist of $\dima$ Hermitian operators ${1, \alher_2 \ldots \alher_{\dima}}$. Note that even if we are originally give a non-Hermitian basis of operators, given any pair of operators, $\al, \al^{\dagger}$,  we can always transform to a Hermitian basis by taking the two combinations ${1 \over 2} (\al + \al^{\dagger})$ and ${i \over 2} (\al - \al^{\dagger})$. Now,  since the basis of operators is Hermitian, the matrix of correlation functions
\be
g_{i j} = \st(\alher_i \alher_j),
\ee
can be taken to be a {\em random} Hermitian positive matrix. Since the basis is Hermitian, the modular operator is just the transpose of this matrix,
\be
(\modop{\st})_{i j} = g_{j i}.
\ee
Similarly, we can take
\be
(\relmod{\st}{\sst})_{i j} = \sst(\alher_j \alher_i),
\ee
to be another random Hermitian positive matrix. 

We can choose to probe this system with a subset of these operators, $\blset \subseteq  \alset $, with $\text{dim}(\blset) = \dima'$. For each subset of $\alset$ we can consider
\be
\spect(\combop_{\blset}) = \spect\left((P_{\blset} \modop{\st} P_{\blset})^{-1}  \cdot P_{\blset} \relmod{\st}{\sst} P_{\blset}\right).
\ee
The largest and smallest eigenvalues in this spectrum give us $||\combop_{\blset}||$ and $1/||\combop_{\blset}^{-1}||$ and we can use these two to define the normed-entropy and the $\chi$-distance when the system is probed with $\blset$. In our example, we take the subset of operators, $\blset$, to be just the first $\dima'$ operators of $\alset$ but it is easy to generalize this to random subspaces of $\alset$.

In figure \ref{figqimat}, we show the normed entropy and the $\chi$-distance starting with 5 different values of $\dima$: (100, 200, 300, 400, 500) and then choosing subsets of operators with all values of $\dima'$ from $2 \ldots \dima$. It is clear that both measures are monotonic as we expect.

\subsection{Separable and entangled states \label{appsepent}}
In the text, our measures of entanglement were given in terms of the distance from the closest separable state. However, given a state, it is not easy to determine whether it is separable or not. 

However, it is often possible to look for an {\em entanglement witness} that can distinguish an entangled state from a separable one \cite{horodecki2001separability}. An entanglement witness is just an operator that has positive expectation values in all separable states; therefore if we find a state where this witness has a negative expectation we know that the state is entangled.  One such entanglement witness, which is commonly used for low-dimensional systems is the partial transpose as we review below \cite{peres1996separability}.

We consider a direct product splitting of $\alset = \alset_1 \otimes \alset_2$ with $\text{dim}(\alset_1) = \dima_1$ and $\text{dim}(\alset_2) = \dima_2$. If a state is separable as in \eqref{sepstatedef} then the matrix of correlations defined in \eqref{twopointfn} can be written as
\be
\label{twoptdecom}
g = \sum_{i} \lambda_i g^{1 i} \otimes g^{2 i},
\ee
where $g^{1 i}$ is a matrix of correlations for elements of $\alset_1$ and $g^{2 i}$ is a matrix of correlations for $\alset_2$ and we have suppressed tensor indices to lighten the notation. As we explained above, this correlation matrix must be Hermitian and positive. 

Now, consider a positive map $\Lambda$ that acts on $\dima_2 \times \dima_2$ matrices and lift its action to $\dima \times \dima$ matrices by considering the map $1 \otimes \Lambda$. When acting on a convex decomposition as in \eqref{twoptdecom}, this map produces a positive matrix.
\be
(1 \otimes \Lambda)(g)  = \sum_i \lambda_i g^{1 i} \otimes \Lambda(g^{2 i}),
\ee
which is a sum of manifestly positive matrices.

However, if $\Lambda$ is {\em not a completely positive map} then if $g$ does not have a decomposition as in \eqref{twoptdecom}, then it is not necessary that $(1 \otimes \Lambda)(g)$ will be a positive matrix. 

If we take $\Lambda$ to be the transpose operation, then this gives us an example of a map between matrices that is positive but not completely positive. The action of $1 \otimes \Lambda$ on a matrix is then given by the ``partial transpose''. Now, if we can find a matrix of correlations, $g$, that has the property that its partial transpose has a negative eigenvalue then this proves that the matrix cannot be written in terms of a convex sum as in \eqref{twoptdecom}.

It is easy to construct a concrete example. The simplest example requires $\dima_1 = \dima_2 = 3$ and we take 
\be
\begin{split}
&\alset_1 = \text{span~of}\{1, \al_1, \al_1^{\dagger} \}; \quad \alset_2 = \text{span~of}\{1, \al_2, \al_2^{\dagger} \}; \\
&\alset = \text{span~of}\{1, \al_1, \al_1^{\dagger}, \al_2, \al_2^{\dagger}, \al_1 \al_2, \al_1 \al_2^{\dagger}, \al_1^{\dagger} \al_2, \al_1^{\dagger} \al_2^{\dagger}\}.
\end{split}
\ee
 Then we only need to find a matrix of two-point correlations that has the property that its partial transpose has at least one negative eigenvalue. 

Consider the following numerically generated
9 dimensional matrix of two-point functions
\be
g_{i j} = \left(
\begin{array}{ccccccccc}
 1. & 0.2 & -0.1 & -0.4 & 0. & 0.4 & 0. & 0.1 & -0.1 \\
 0.2 & 1. &  0.1 i & 0. & -0.4 &  0.1 i & 0.1 & 0. &  0.3 i \\
 -0.1 &  -0.1 i & 1. & 0.4 &  -0.1 i & -0.4 & -0.1 &  -0.3 i & 0. \\
 -0.4 & 0. & 0.4 & 1. & 0.2 & -0.1 &  0.3 i &  -0.1 i &  0.2 i \\
 0. & -0.4 &  0.1 i & 0.2 & 1. &  0.1 i &  -0.1 i &  0.3 i & -0.4 \\
 0.4 &  -0.1 i & -0.4 & -0.1 &  -0.1 i & 1. &  0.2 i & 0.4 &  0.3 i \\
 0. & 0.1 & -0.1 &  -0.3 i &  0.1 i &  -0.2 i & 1. & 0.2 & -0.1 \\
 0.1 & 0. &  0.3 i &  0.1 i &  -0.3 i & 0.4 & 0.2 & 1. &  0.1 i \\
 -0.1 &  -0.3 i & 0. &  -0.2 i & -0.4 &  -0.3 i & -0.1 &  -0.1 i & 1. \\
\end{array}
\right).
\ee
We can check that the spectrum of this matrix and its partial transpose are given by
\be
\begin{split}
&\spect{\left(g_{i j}\right)} \approx \{2.10,1.78,1.37,1.19,1.07,0.78,0.38,0.19,0.15\},\\
&\spect{\left(g_{i j}^{\text{pt}}\right)} \approx \{2.06,1.78,1.41,1.21,0.99,0.76,0.59,0.23,-0.02\},
\end{split}
\ee
and we see that the last eigenvalue of the partial transpose is negative. So this matrix of correlations represents an {\em entangled} state.

While we applied the partial transpose criterion to the matrix of two-point functions, note that we could just as well have applied it to the modular operator. Second, we should caution the reader that the matrix of two-point functions is quite different from the density matrix itself, even though both these matrices are positive. For instance, in the example above, since the operators in $\alset_1$ and $\alset_2$ commute, the matrix of two-point functions reflects this symmetry and not every positive $9 \times 9$ matrix is allowed. One such constraint, in the basis above, is that we must have $g_{2 4} = g_{42}$ because  $\st(\al_1 \al_2) = \st(\al_2 \al_1)$.

\bibliographystyle{JHEP}
\bibliography{references}

\end{document}